\documentclass[referee]{aa} % for a referee version
\usepackage{graphicx}
\usepackage{txfonts}
\usepackage{url}
\usepackage{natbib}

\bibpunct{(}{)}{;}{a}{}{,}  % to follow the A&A style
\newcommand{\HI}{\ion{H}{I}}  
\newcommand{\kms}{\hbox{km\,s$^{-1}$}}
\newcommand{\degree}{$^{\circ}$}
\newcommand{\nh}{$N_{\rm HI}$}

\newcommand{\slvalueNEP}{0.097}
\newcommand{\slvalue}{0.0981}
\newcommand{\labvsgbtratio}{1.0288}
\newcommand{\aragodirection}{\hbox{$11.879\degr$}} %11.878953 measured in
\newcommand{\subreflectordirection}{\hbox{$12.3\degr$}} % 17.9 - 5.6
\newcommand{\spilloverdirection}{\hbox{$12\degr$}} % fix this to be center
\newcommand{\rigging}{\hbox{$50\fdg3$}} % the rigging angle
\newcommand{\tilt}{\hbox{$17\fdg0$}} % the subreflector tilt angle

\begin{document}
   \title{Accurate Galactic 21-cm \HI\ measurements\\
 with the NRAO Green Bank Telescope}

%   \subtitle{No subtitle}

   \author{A. I. Boothroyd\inst{1}
          \and
          K. Blagrave\inst{1}
          \and
          Felix J. Lockman\inst{2}
          \and
          P. G. Martin\inst{1}
          \and
          D. Pinheiro Gon\c{c}alves\inst{3}
          \and
          Sivasankaran Srikanth\inst{4}
          }

   \institute{Canadian Institute for Theoretical Astrophysics, University of Toronto\\
              60 St.\ George Street, Toronto, Ontario, Canada\ \ M5S 3H8\\
              \email{boothroy@cita.utoronto.ca}
              \email{blagrave@cita.utoronto.ca}
              \email{pgmartin@cita.utoronto.ca}
         \and
             National Radio Astronomy Observatory \\
             P.O. Box 2, Green Bank, WV USA \ 24944\\
             \email{jlockman@nrao.edu}
         \and
              Department of Astronomy and Astrophysics, University of Toronto \\
              50 St.\ George Street, Toronto, Ontario, Canada\ \ M5S 3H4\\
             \email{goncalves@astro.utoronto.ca}
         \and
              NRAO Technology Center \\
             1180 Boxwood Estate Rd., Charlottesville, VA 22901, USA \\
             \email{ssrikant@nrao.edu}
             }
\date{Received 8 July 2011 /
 Accepted 6 Oct 2011}

% \abstract{}{}{}{}{} 
% 5 {} token are mandatory, but sometimes contents are optional
 
  \abstract
   {}
{We devise a data reduction and calibration system for producing
  highly-accurate 21-cm \HI\ spectra from the Green Bank Telescope (GBT)
  of the NRAO.}
{A theoretical analysis of the all-sky response of the GBT at 21~cm is
  made, augmented by extensive maps of the far sidelobes.  Observations
  of radio sources and the Moon are made to check the resulting aperture
  and main beam efficiencies.}
% results heading (mandatory} 
{The all-sky model made for the response of the GBT at 21~cm is used
  to correct for ``stray'' 21-cm radiation reaching the receiver
  through the sidelobes rather than the main beam.  This reduces
  systematic errors in 21-cm measurements by about an order of
  magnitude, allowing accurate 21-cm \HI\ spectra to be made at about
  9\arcmin\ angular resolution with the GBT.  At this resolution the
  procedures discussed here allow for measurement of total integrated
  Galactic \HI\ line emission, $W$, with errors of 3~K~\kms,
  equivalent to errors in optically thin \nh\ of
  $5\times10^{18}$~cm$^{-2}$.}
   {}


   \authorrunning{Boothroyd {\it et al.}}

   \titlerunning{Accurate GBT \HI\ Measurements}

   \maketitle

\section{Introduction} \label{intro}

Accurate spectra of Galactic \HI\ are needed for many areas of
research, from studies of interstellar gas and dust
\citep[e.g.,][]{1981Heiles,1988Boulanger,boul96,2005LockmanCondon,Kalberla2009,Abergel2011},
to determination of the distribution and abundances in interstellar
gas \citep[e.g.,][]{Hobbs1982,Albert1993,Shull2009}, to correction of
extragalactic radiation for absorption by the ISM and removal of
foregrounds
\citep[e.g.,][]{Jahoda1985,Hasinger1993,Snowden1994,Hauser1998,Puccetti2011,Ade2011}.
Although the highest resolution \HI\ observations are obtained with
aperture synthesis techniques, Galactic \HI\ emission is smoothly
distributed across the sky, i.e., most of the power is in the lowest
spatial frequencies, and so filled aperture (i.e., single dish)
observations are essential for determining accurate \HI\ spectra
\citep[e.g.,][]{1976Heiles,Green1993,Dickey2001,Kalberla2009}.  The
most accurate single-dish Galactic \HI\ measurements are now capable
of determining $W$, the integrated emission over the line profile,
with an error of just a few percent, and hence yield \nh, the total
column density of Galactic \HI\ to the same precision provided that
opacity effects are small \citep{2011Wakker}.

The Robert C.\ Byrd Green Bank Telescope (GBT) is a 100-m diameter
filled-aperture radio telescope that has been used for many studies of
Galactic \HI\ and the extended \HI\ emission around nearby galaxies
\citep[e.g.,][]{2011Hunter}.  In this paper we describe techniques
whereby we are able to produce high-quality 21-cm spectra with this
instrument with overall calibration errors of only a few percent. For
illustration, we consider data from recent \HI\ surveys undertaken to
study the gas-dust correlation and dynamics of high latitude cirrus
\citep{Blagrave2010,Abergel2011,Martin2011}, although the methods
described here can be applied to any 21-cm \HI\ data taken with the
GBT.
It is important to note that measurement of accurate extragalactic
\HI\ is much more straightforward than for Galactic \HI\ as there is
little confusing emission entering through sidelobes at $|v| \gtrsim
200$ km s$^{-1}$; errors in extragalactic \HI\ profiles can be as
small as $3\%$, arising largely from instrumental baseline effects
\citep[e.g.,][]{Hogg2007}.

In Sect.~\ref{GBTinstrumentation} we describe the telescope and its
instrumentation.  Section~\ref{HIobsTechnique} considers the observing
techniques used for our Galactic \HI\ measurements.  In
Sect.~\ref{dataReduction} we describe the steps taken to reduce and
calibrate the data.  Section~\ref{aperturebeamefficiency} describes a
theoretical calculation of the main beam properties, the calibration
of the antenna temperature scale, conversion to brightness
temperature, and several independent cross-checks of the accuracy of
these steps.  In Sect.~\ref{sidelobes} measurements of the antenna
response up to $60\degr$ from the main beam are described, leading to
an all-sky antenna response suitable for correcting for ``stray
radiation'' detected through the sidelobes rather than the main beam.
Section~\ref{examples} illustrates the effects of the correction on
several 21~cm spectra.  In Sect.~\ref{testerrors} several tests of the
data reduction method and our estimates of the errors are documented.
In Sect.~\ref{absolutecalibration} we discuss the absolute
calibration.  Section~\ref{conclusions} presents a summary and
conclusions.

\section{The Green Bank Telescope and its instrumentation}
 \label{GBTinstrumentation}

\begin{figure}
\centering
\includegraphics[angle=-90,width=0.9\linewidth]{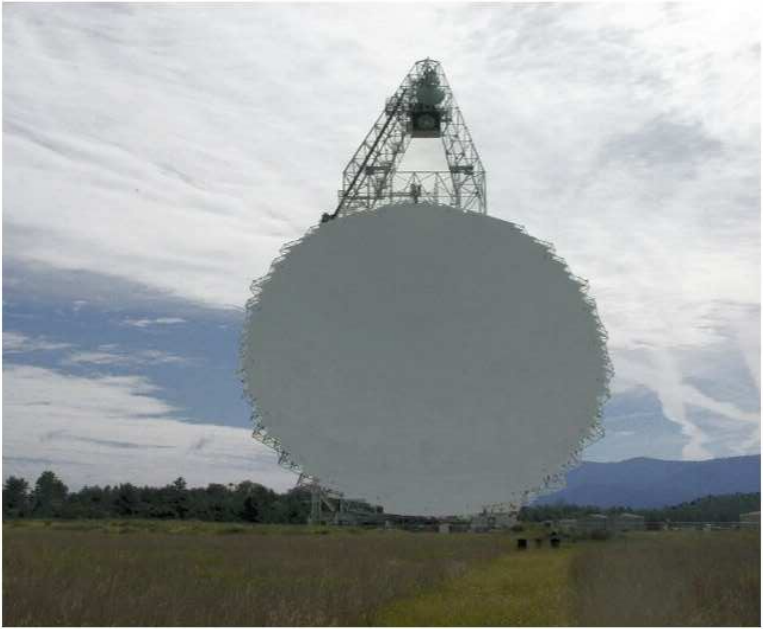}
\caption{The Green Bank Telescope in a view that shows its unblocked
  100-meter diameter aperture.}
\label{GBT_front}
\end{figure}

\begin{figure}
\centering
\includegraphics[angle=0,width=0.8\linewidth]{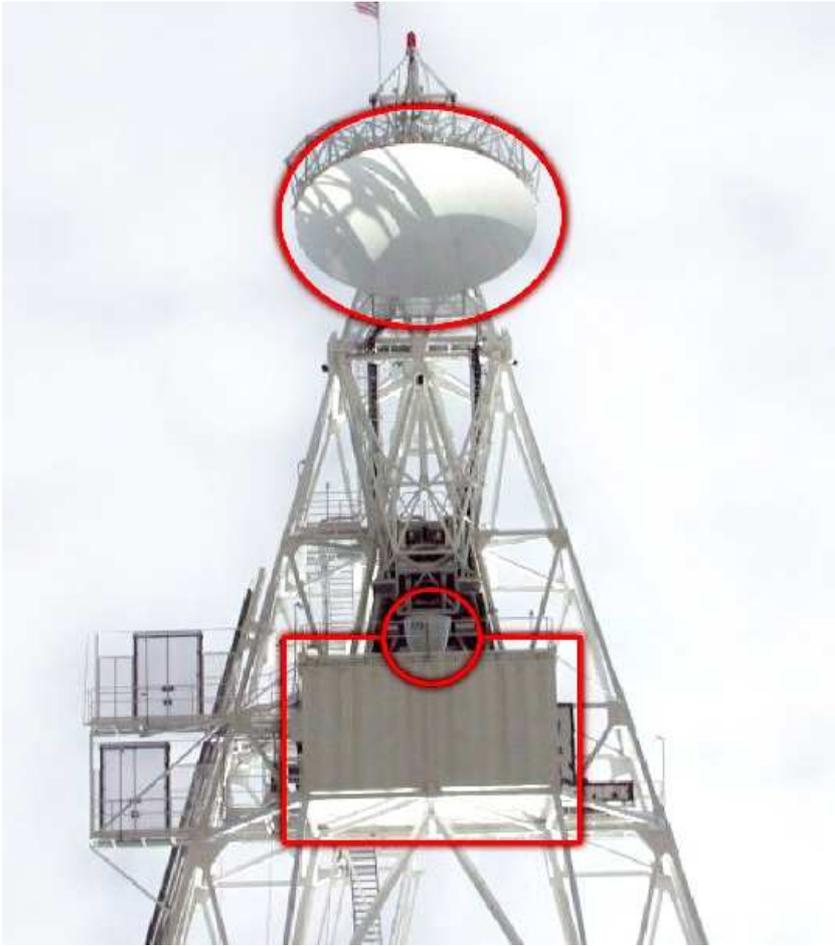}
\caption{The focal area of the GBT as seen from the surface of the main
  reflector, showing outlined in red, bottom to top, (i) the receiver
  room, with (ii) the L-band feed horn pointing upward at (iii) the
  8-meter subreflector.  Just below the lower edge of the subreflector
  the lower portion of a screen is visible.  This functions to direct
  feed spillover radiation that would otherwise strike the telescope arm
  back into the main reflector and onto the sky.}
\label{GBT_arm}
\end{figure}

\begin{figure}
\centering
\includegraphics[angle=0,width=0.6\linewidth]{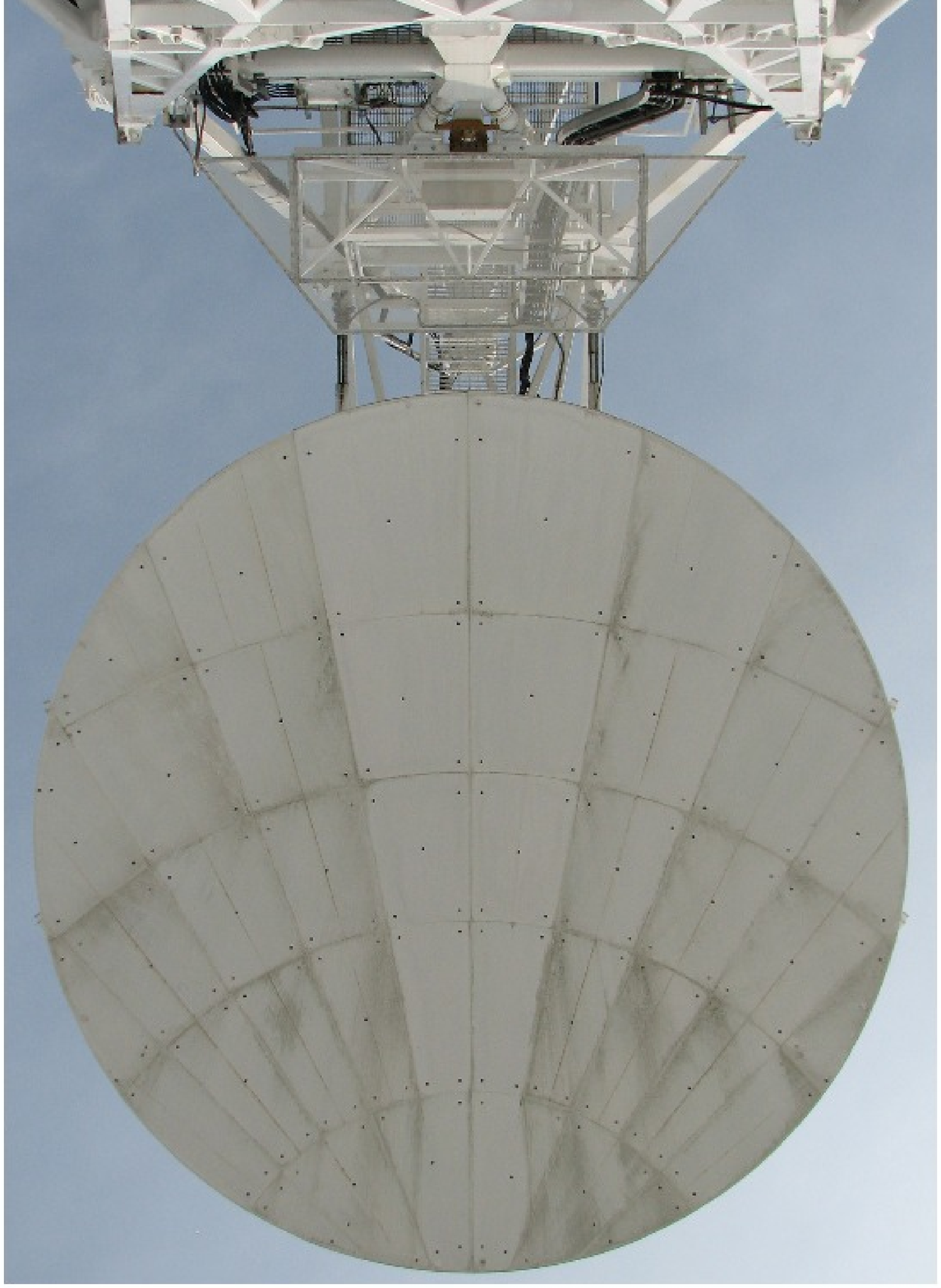}
\caption{The GBT subreflector and feed arm seen from the secondary focal
  point atop the receiver room.  The screen redirects feed spillover
  away from the arm and down into the main reflector.
}
\label{GBT_sub_screen}
\end{figure}

The GBT is a 100-m diameter dual offset Gregorian reflector with a
large, unblocked aperture on an azimuth-elevation mount
(Fig.~\ref{GBT_front}, Fig.~\ref{GBT_arm}, Fig.~\ref{GBT_sub_screen};
\citealp{Prestage09}). The surface consists of 2004 panels mounted on
motor-driven actuators capable of real-time adjustment to compensate
for gravitational astigmatism and other surface distortions allowing
it to achieve a surface rms $<250\mu$.  We used the telescope with the
surface in passive mode where it has a typical rms error $\approx900
\mu$, and thus at
21~cm (1420~MHz) a typical surface efficiency of 0.997.  
Because of gravitational distortions and thermal effects,
the passive surface rms accuracy of the GBT is expected to vary
between $500\mu$ and $1200\mu$ at the extremes.  This would cause a
slight change in the main beam shape and a point-source gain change
with a range of up to $0.4\%$.  The effect on \HI\ observations is
more complex and likely less important as it depends on the
convolution of the detailed beam shape with the \HI\ sky brightness.
The main beam
has a FWHM that is $9.1\arcmin \times 9.0 \arcmin$ in the
cross-elevation and elevation directions, respectively.  At 21~cm, the
aperture efficiency $\eta_{\rm a} = 0.65$ and the main beam efficiency
$\eta_{\rm mb} = 0.88$.  The derivation of these quantities will be
discussed in Sect.~\ref{calcside} and Appendix~\ref{calcSri}.

The 21-cm ``L-band'' receiver used for these measurements is located
at the secondary Gregorian focus and illuminates an 8-m diameter
subreflector with an average edge taper of $-14.7$~dB
\citep{Srikanth1993}.  The receiver is cryogenically cooled, accepts
dual linear polarization, and has a total system temperature, $T_{\rm
  sys}$, at the zenith of 18~K. A diode injects a fixed amount of
noise into the waveguide just after the feed horn, before the
polarizer and amplifier, and can be modulated rapidly for calibration.
This was used to establish a preliminary scale for the antenna
temperature, $T_{\rm a}$.  These noise sources are typically very
stable; we see no evidence for variation of the L-band receiver
calibration diode over several years (Sect.~\ref{testerrors}).

\section{\HI\ 21-cm line data acquisition} \label{HIobsTechnique}

\subsection{Mapping}

The \HI\ 21-cm observations discussed here were made with On-the-Fly
(OTF) mapping: ``scanning'' or moving the telescope in one direction,
typically Galactic longitude or Right Ascension, while taking data
continuously.  The integration time and telescope scan rate must be
chosen so that samples are taken no more coarsely than at the Nyquist
interval, $\approx {\rm FWHM}/2.4 = 3.8\arcmin$ for the GBT at 21~cm,
and ideally at half that interval to avoid beam broadening in the
scanning direction \citep{Mangum07}.  Areas were mapped by stepping
the scans in the fixed coordinate (the cross-scan direction) and
reversing the scan direction.  In the \HI\ surveys discussed here
\citep{Martin2011} scans up to $5{}\degr$ long were made, but for
practical purposes larger regions were broken up into smaller areas
with dimensions between $2{}\degr \times 2{}\degr$ and $4{}\degr
\times 4{}\degr$ that were mapped independently.

\subsection{Spectrometer setup}\label{specsetup}

The GBT autocorrelation spectrometer as configured for these
measurements has 16k~channels over a 12.5~MHz band with nine-level
sampling in each linear polarization, for a velocity resolution of
0.16~\kms\ in the 21-cm line.  Early in the data reduction procedure
the resolution was reduced to 0.80~\kms\ by filtering the spectra with
an eleven-channel Hanning smoothing function and resampling every
fifth channel.  This maintains the independence of each channel, while
minimizing aliasing effects.

The local oscillator was modulated to move the center of the
spectrometer band between a ``signal'' and ``reference'' frequency,
separated by 2.5~MHz (528~\kms).  The noise source was modulated
synchronously with the frequency switching to calibrate the receiver
gain at both signal and reference frequencies every second.  Because
the modulated separation is much smaller than the 12.5~MHz covered
instantaneously by the spectrometer, the emission in the 21-cm line
was always being observed.  This ``in-band'' frequency switching gives
a factor of two increase in observing speed over ``out-of-band''
frequency switching, and an rms noise in antenna temperature,
$\sigma_{\rm Ta}$, of approximately 0.25~K in a 1~\kms\ channel in
1~second for the average of the two polarizations.  For the basic
integration time for the survey data, 4~seconds, and velocity
resolution 0.80~\kms, the resulting rms noise is 0.16~K for
emission-free channels (see Sect.\ref{specnoise}).  Some regions were
measured several times to check the reproducibility and to improve the
sensitivity.

\subsection{Spectral baselines}\label{baselines}

\begin{figure}
\centering
\includegraphics[angle=90,width=0.85\linewidth]{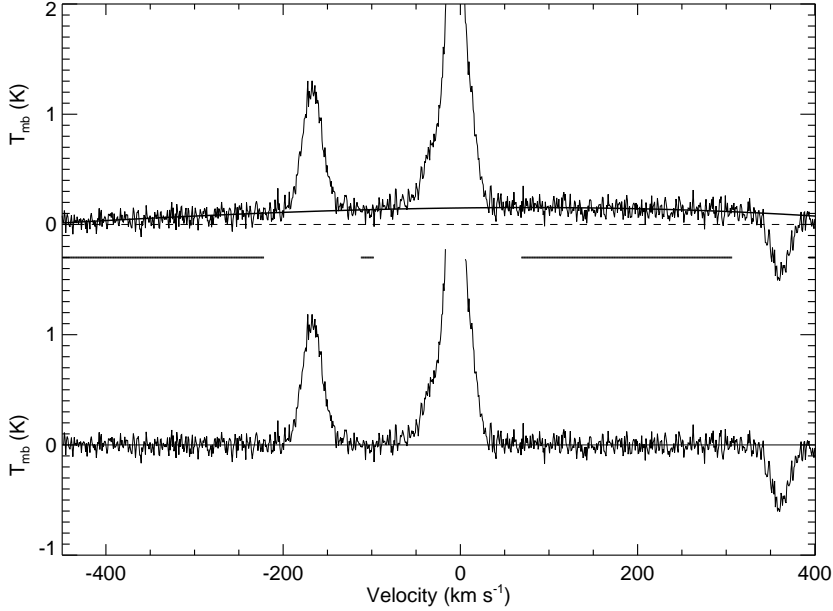}
\caption{
Brightness temperature spectrum from the North Ecliptic Pole (NEP)
data cube at $(l,b) = (90\fdg318, 34\fdg432)$, illustrating aspects of
the baseline removal.  The negative feature at 370~\kms\ results from
the 2.5~MHz frequency-switching: it is an offset half-amplitude
inversion of the $-170$~\kms\ feature.
The upper panel shows a typical third-order polynomial baseline fit to
the central 4~MHz of the frequency-switched spectrum; the horizontal
line segments mark the ``emission-free'' channels determined
iteratively to be part of the baseline (see text).
The lower panel displays the spectrum following baseline removal.  }
\label{NEPbaseline}
\end{figure}

The frequency-switched spectra produced by the GBT spectrometer (e.g.,
Fig.~\ref{NEPbaseline}) have remarkably flat instrumental baselines
over the central 4~MHz, even more so for the XX compared to the YY
polarization.  Nevertheless, an important step in our GBT data
reduction procedure (Sect.~\ref{dataReduction}) is removal of any
residual instrumental baseline.  As is usual, the baseline is
approximated by a low-order polynomial whose coefficients are fit by
linear least-squares to the ``emission-free'' channels of each
spectrum.
For the GBT spectra we found that over the central $\sim4$~MHz ($-450
< v < +400$ \kms) a third-order polynomial was adequate, the next
order term being not statistically justified.  Except for diagnostic
purposes, we fit baselines to spectra after correction for stray
radiation and assembly into a data cube (Sect.~\ref{procedure}).  In a
very few spectra, out-of-band interference results in unsalvageable
spectra with high-order polynomial baselines.  These are easily
identified by eye, and in most cases we were able to reobserve these
positions.  Alternatively, these spectra are flagged for exclusion
from the map-making process.

For our mapping observations with the GBT, the instrumental baseline
was found to change slowly with time, so that the coefficients of the
polynomial fit are highly correlated between spectra along a scan. For
noisy spectra there are potential advantages to fitting a baseline to
the average of sequential spectra \citep{Lockman1986}, but we did not
implement this as fitting of individual polynomials gave adequate
results.  It is possible to take advantage of an iterative baseline
fitting technique.  This brings in more channels than one normally
obtains from a conservative estimate of the location of the
emission-free end channels for a particular map, even locating
emission-free parts of the spectrum between emission features.

We fit baselines following the iterative technique used by
\citet{1994Hartmann} for the Leiden/Dwingeloo Survey.  Prior to any
fitting, each spectrum is smoothed by a 20-channel (roughly 16~\kms)
boxcar to accentuate real velocity features.  We used Hartmann's
definition of a ``velocity feature'' present in a residual spectrum
from which an estimated baseline has been removed.  These are found by
first identifying a significant ($4.0\sigma$) positive peak and then
adding neighbouring channels until the residual goes negative; ten
additional channels (about 8~\kms) at both ends of each velocity
feature are also flagged for omission.  Following the identification
of all such features, there remain the ``emission-free'' baseline
channels to be used in the next fit.  The smoothed baseline spectrum
is fit successively with a series of polynomials monotonically
increasing from linear to third-order, after every iteration
subtracting this updated baseline and identifying and flagging new
velocity features with significant residual peaks.  Because the
spectra are frequency-switched, any velocity emission feature will
show up as a half-amplitude inverted feature offset by the switching
frequency, 2.5~MHz.  These channels are also flagged for exclusion.
Finally, a third-order polynomial is fit to the remaining list of
emission-free channels in the {\it original} spectrum and subtracted.
Usually $\sim 600$ emission-free channels are used for the fit.
Figure~\ref{NEPbaseline} illustrates a typical result of the iterative
baseline fitting process, for a typical instrumental baseline.

Occasionally we found \HI\ emission from background galaxies in our
spectra, usually in the end channels beyond the Galactic emission, but
sometimes even overlapping it.  This makes determining the
instrumental baseline challenging.  While this can be treated on a
case-by-case basis, these pixels are simply masked as unsuitable for
analysis of Galactic \HI.

\subsection{Treatment of radio frequency interference (RFI)}

The main source of RFI in these data is an oscillator in the GBT
receiver room which can produce narrow-band spurious signals in the
data.  These are extremely stable and $<<1$ kHz in width. They can be
identified easily and removed.  A series of eight frequency ranges in
which RFI had been seen were systematically inspected by averaging
many spectra together in the topocentric velocity frame of reference.
Any $3.5\sigma$ positive deviations from the median over the suspect
frequency range were flagged.  Finally data in a group of five
channels around the flagged frequency were replaced with values from a
linear interpolation of the surrounding channels.  This channel
replacement is done at the highest velocity resolution (0.16~\kms) on
the four observed spectrometer phases (sig/calon, sig/caloff,
ref/calon, ref/caloff) prior to producing the final calibrated
frequency-switched spectra and prior to any subsequent spectral
smoothing.  Efforts are underway to replace the interfering device.

\section{Data reduction} \label{dataReduction}

The measured quantity, $T_{\rm a},$ is related to the 21-cm \HI\ sky
brightness $T_{\rm b}$ by

\begin{equation}
T_{\mathrm{a}} = \eta_{\mathrm{r}} \int_{4{\mathrm{\pi}}}
P(\theta,\phi) T_{\mathrm{b}}(\theta,\phi)
{\mathrm{e}}^{-\tau_{\mathrm{a}}} \mathrm{d}\Omega
\label{basicTa}
\end{equation}
where ${\rm e}^{-\tau_{\rm a}}$ is the direction-dependent atmospheric
extinction and $P$ is the antenna power pattern; the integration is
over the entire $4{\rm \pi}$~sr and this ``all-sky'' integral of $P$
is unity (later we also refer to the antenna gain $G = 4{\rm \pi} P$
relative to isotropic, where by definition $G_{\rm
  isotropic}(\theta,\phi) = 1$).  The quantity $\eta_{\rm r}$ accounts
for resistive losses in the system which are less than $1\%$.  Our
procedure in Sect.~\ref{gettingTa} for calibration of $T_{\rm a}$
produces an antenna temperature scale that is independent of
$\eta_{\rm r}$ and we will thus drop this factor from subsequent
equations.  Equation~\ref{basicTa} can be separated into terms that
come from the main beam and from elsewhere on the sky:

\begin{equation}
T_{\mathrm{a}} =
\mathrm{e}^{-\tau_{\mathrm{mb}}}\eta_{\mathrm{mb}}\langle
T_{\mathrm{mb}} \rangle + \int_{\Omega_{\mathrm{sl}}} P(\theta,\phi)
T_{\mathrm{b}}(\theta,\phi)\mathrm{e}^{-\tau_{\mathrm{a}}}
\mathrm{d}\Omega
\label{eq-2}
\end{equation}
where ${\rm e}^{-\tau_{\rm mb}}$ is the atmospheric extinction in the
direction of the main beam, $\eta_{\rm mb}$, the main beam efficiency,
is the fraction of the total power accounted for by the main beam over
$\Omega_{\rm mb}$, $\Omega_{\rm sl}$ is the area of the sky outside
the main beam ($\Omega_{\rm sl}=4{\rm \pi}-\Omega_{\rm mb}$).  The
desired quantity to be measured, $\langle T_{\rm mb} \rangle$, is the
\HI\ brightness temperature averaged over the main beam:

\begin{equation}
\langle T_{\mathrm{mb}} \rangle = {1\over{\Omega_{\mathrm{mb}}}}
\int_{\Omega_{\mathrm{mb}}} T_{\mathrm{b}}(\theta,\phi)
\mathrm{d}\Omega_{\mathrm{mb}}. 
\label{avg-Tb-def}
\end{equation}

While $T_{\rm b}$ and $T_{\rm a}$ are functions of frequency because
of Doppler shift, it is assumed that the other quantities are constant
over the relatively narrow frequency range of Galactic HI emission.
Equation~\ref{eq-2} can thus be written

\begin{equation}
\langle T_{\mathrm{mb}} \rangle =
{\mathrm{e}^{\tau_{\mathrm{mb}}}\over{\eta_{\mathrm{mb}}}} \left[
T_{\mathrm{a}} - \int_{\Omega_{\mathrm{sl}}} P(\theta,\phi)
T_{\mathrm{b}}(\theta,\phi)\mathrm{e}^{-\tau_{\mathrm a}}
\mathrm{d}\Omega \right].
\label{removestray}
\end{equation}

The aperture efficiency $\eta_{\rm a}$ and beam efficiency $\eta_{\rm
  mb}$ are not necessarily independent \citep[e.g.,][]{Goldsmith2002}.
For a dish of diameter, $D$, at wavelength, $\lambda$, with a Gaussian
main beam, $\theta_{\rm FWHM}$, and with typical values for the main
reflector edge taper, we have
$\eta_{mb} \equiv \Omega_{\mathrm{mb}}/\Omega_{\mathrm{a}}$, 
$\Omega_{\mathrm{a}} = 4 \lambda^2/(\eta_{\mathrm{a}} \pi D^2)$,
$\Omega_{\mathrm{mb}} \approx 1.13 \theta_{\mathrm{FWHM}}^2$, and
$\theta_{\mathrm{FWHM}} \approx 1.25 \lambda/D$.
Thus, $\eta_{\rm mb} \approx 1.4 \eta_{\rm a}$ for the GBT at 21~cm.
However, this relationship is not necessarily precise for real
systems, and so we treat the two terms as independent quantities that
have to be derived and checked separately.

For simplicity, the observable $\langle T_{\rm mb} \rangle$ will be
referred to as $T_{\rm mb}$ or simply $T$ from henceforth.

\subsection{Stray radiation} \label{strayintro}

The second term in Eq.~\ref{removestray} accounts for \HI\ emission
that enters the receiver through the sidelobes of the telescope rather
than through the main beam. This is called ``stray'' radiation.  Just
as the main beam efficiency, $\eta_{\rm mb}$, reflects the fraction of
the power pattern in the main beam, we define a sidelobe efficiency,
$\eta_{\rm sl}$,

\begin{equation}
\eta_{\mathrm{sl}} \equiv \int_{\Omega_{\mathrm{sl}}} P(\theta,\phi)
\mathrm{d}\Omega = 1 - \eta_{\mathrm{mb}}.
\end{equation} 

The unblocked aperture of the GBT eliminates the scattering sidelobes
that plague most other radio telescopes, but at 21~cm the relatively
low edge taper of the feed (originating from mechanical limitations on
its size and weight, see \citealp{Norrod1996}) results in a
``spillover lobe'' past the secondary reflector.
Because there is 21-cm \HI\ emission from all directions of the sky at
a level \nh\ $\gtrsim 4 \times 10^{19}$ cm$^{-2}$
\citep{Lockman1986,Jahoda1990}, there will be a contribution to every
spectrum from \HI\ emission entering through the sidelobes, and so the
``stray'' radiation spectrum must be calculated and removed as a part
of data reduction and calibration.  Prior to construction of the GBT
the only unblocked radio telescope useful for 21-cm measurements was
the Bell Labs horn-reflector which had a main beam size $\approx
3\degr \times 2\degr$ and a spillover lobe as well
\citep{1972Wrixon,1992Kuntz}

Stray radiation originating from large angles is Doppler-shifted with
respect to the direction of the main beam, so that the stray radiation
spectrum correctly shifted to the local standard of rest (LSR) depends
not only on the location of the sidelobe on the sky but also on the
date and time of the observation \citep{Kalberla2005}.  For an Alt-Az
telescope like the GBT, the sidelobe pattern, which is fixed in
telescope coordinates, also rotates on the sky about the main beam as
the Local Sidereal Time (LST) changes.  A direction observed at
varying LST will have a varying component of stray radiation.

Stray radiation has been an issue for Galactic 21-cm science for
50~years \citep{Vanwoerden1962,1976Heiles} and many groups have
devised methods to suppress or remove it
\citep[e.g.,][]{1972Wrixon,Kalberla1980,Lockman1986,Hartmann1996,Higgs2005,Kalberla2010}.
The most accurate methods solve Eq.~\ref{removestray} for $\langle
T_{\rm mb} \rangle$ using models of the antenna power pattern,
$P$, and all-sky $T_{\rm b}(\theta,\phi)$ maps.

Because the most important sidelobes of the GBT cover large areas of
the sky, relatively low angular resolution 21-cm surveys can be used
for the $T_{\rm b}(\theta,\phi)$ term in estimating the stray
component, e.g., the Leiden/Argentine/Bonn (LAB) survey
\citep{Kalberla2005}.  However, this is not true for the first few
diffraction sidelobes that lie close to the main beam.  These have
angular structure on scales like that of the main beam and would
require knowledge of the \HI\ sky at that level of detail for their
removal.  For the GBT, however, the near sidelobes are quite small and
contain $<1\%$ of the telescope response (Sect.~\ref{calcside}).  We
do not correct for these, in effect assuming that their small
component of the telescope's response samples nearly the same emission
as the main beam.

\subsection{Procedure} \label{procedure}

The data reduction procedure thus involves 
calibration of the intensity scale to antenna temperature, 
calculation and subtraction of the stray radiation spectrum,
and correction for the main beam efficiency and
the atmosphere.  Data reduction additionally requires 
removal of an instrumental baseline, and for maps, 
interpolation of the sampled spectra into a data cube.  The latter two
steps are sometimes interchanged; the interpolated spectra, especially
if combining several observations of a region, are less noisy so that
baseline removal at that final stage might be preferable.

The initial calibration of the data to an approximate $T_{\rm a}$
antenna temperature scale used values for the receiver calibration
noise source that were determined by measurements in the laboratory
for each receiver polarization channel.  This part of the data
reduction was performed using the NRAO program GBTIDL.  A constant
calibration temperature was assumed over the 12.5-MHz band.  We
checked this part of the calibration, and we correct it by a small
amount using our own measurements and calculations as described below
in Sect.~\ref{gettingTa}.

Our implementation of the stray radiation correction is described in
Sect.~\ref{implementstray} and Appendix~\ref{program}, making use of
the sidelobe pattern established in Sect.~\ref{sunscans}.

Calculation of the main beam efficiency is described in
Sect.~\ref{calcside}.
The NRAO data reduction program GBTIDL will perform an approximate
correction for atmospheric extinction\footnote{Jim Braatz, October 30,
  2009: Calibration of GBT Spectral Line Data in GBTIDL v2.1, from
  http://www.gb.nrao.edu/GBT/DA/gbtidl/gbtidl\_calibration.pdf} but we
chose to make the correction independently, using $\tau_{\rm zenith} =
0.01036 \pm 0.00059$, the weighted mean of the measured values of
\citet{Williams1973} and \citet{vanZee+1997}, with a model for the
atmospheric air mass (Appendix~\ref{program}).

After the spectra were calibrated and corrected for stray radiation,
an instrumental baseline was removed from each spectrum by fitting a
third-order polynomial to emission-free velocities.  It is important
that the stray radiation be removed before baseline fitting, lest weak
stray wings be mistaken for instrumental baseline.
For mapped regions, data cubes were constructed in classic AIPS,
averaging together spectra from the two polarizations, using the
optimal tapered Bessel function for interpolation \citep{Mangum07}.
In our data reduction procedure, baselines were removed after creating
the cubes.  For details see Sect.~\ref{baselines}.

Note that nowhere during this procedure are ``standard'' \HI\ regions,
like S6 \citep{Williams1973} and S8 \citep{Kalberla1982}, used to
calibrate the GBT intensity scale.  Because of the clean optics of the
GBT, we preferred to determine its characteristics from basic
calculations and {\emph {radio continuum}} flux density calibration
sources.  We did, however, observe the above two standard \HI\
directions and we discuss these measurements in Sects.~\ref{S6S8calib}
and \ref{S6S8calibAbs}.  We also compared our spectra to those of the
LAB survey (Sect.~\ref{LABscale}).

\section{Aperture and main beam efficiencies} \label{aperturebeamefficiency}

\subsection{Calculation of the efficiencies} \label{calcside}

\begin{figure}
\centering
\includegraphics[angle=-90,width=0.85\linewidth]{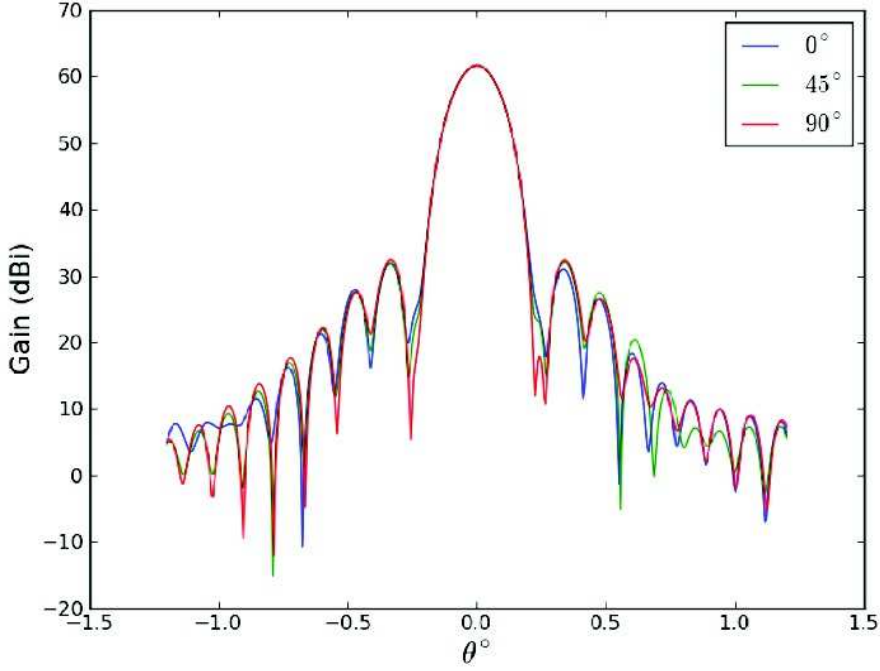}
\caption{GBT antenna gain above isotropic for the main beam and near
  sidelobes at 1.4 GHz, as a function of polar angle $\theta$ from the
  main beam in several planes defined by $\phi$ (see
  Appendix~\ref{calcSri} for the antenna code used for this
  calculation).  The forward gain is 61.6~dBi
  (Table~\ref{eta_a_table}).  }
\label{mainbeam}
\end{figure}

A theoretical estimate of the all-sky response of the GBT was
calculated using a reflector antenna code developed at the Ohio State
University as described in Appendix~\ref{calcSri}.  This incorporates
information on the detailed illumination pattern of the 21-cm feed on
the GBT subreflector as measured after construction of the receiver
\citep{Srikanth1993}. Table~\ref{eta_a_table} shows the calculated
on-axis or forward gain and aperture efficiency as a function of
frequency for the L-band receiver.  The variation in $\eta_{\rm a}$
results from measured changes in the L-band receiver illumination
pattern with frequency. The calculated gain $G(\theta,\phi)$ of the
GBT at 1.4~GHz within $1\fdg2$ of the main beam is shown in
Fig.~\ref{mainbeam} for radial cuts along polar angle $\theta$ in
planes at several angles $\phi$ (see the angle definitions in
Appendix~\ref{calcSri}). The units are dBi (logarithmic units relative
to an isotropic beam) and the value at $\theta = 0$\degree\ is that in
Table~\ref{eta_a_table}.  The first sidelobe is calculated to be 29~dB
below the forward peak of the main beam.  Observations also indicate
that the GBT's main beam is exceptionally clean with near-in sidelobes
all about 30~dB below the main beam gain \citep{Robishaw2009}.

Table~\ref{eta_mb_table} gives the fraction of the total power pattern
lying within a given radius around the main beam, at radii
corresponding to the minima in Fig.~\ref{mainbeam}.  This shows that
87.7\% of the antenna response lies within $0\fdg4$ of the main beam,
increasing by only a small amount to 88.1\% at $1\fdg2$ radius,
consistent with the observations that the near sidelobes are at very
low levels.  We adopt a value $\eta_{\rm mb} = 0.88$.  This value was
also derived independently during the initial calibration of the
L-band receiver on the GBT \citep{Heiles2003}.

We define the ``far'' sidelobes as those arising at angles $\theta
>1\degr$ from the main beam.
Compared to $\eta_{\rm sl} \approx 0.1$, the potential for stray
radiation arising within $0\fdg4-1{}\degr$ of the main beam is
negligible ($\Delta \eta \approx 0.004$); furthermore, the brightness
of this radiation $T_{\rm b}$ will not be grossly different from that
seen on axis, and it will not be Doppler shifted as in the far
sidelobes.

\begin{table}
\caption{Calculated forward gain and aperture efficiency for
  the GBT L-band receiver}
\label{eta_a_table}
\centering
\begin{tabular}{c c c}
\hline \hline
Frequency & Forward Gain  & $\eta_{\rm a}$ \\
(GHz) & (dBi)  \\
\hline
1.15 & 60.1 & 0.708 \\
1.40 & 61.6 & 0.654 \\
1.73 & 63.7 & 0.712 \\
\hline
\end{tabular}
\end{table}

\begin{table}
\caption{Calculated fractional power pattern at 1.4 GHz as a
    function of angle from the beam center}
\label{eta_mb_table}
\centering
\begin{tabular}{c c}
\hline \hline
$\theta$  & $P(\theta)$ \\
\hline
$0\fdg26$ & 0.871  \\
$0\fdg41$ & 0.877  \\
$0\fdg55$ & 0.880  \\
$0\fdg67$ & 0.881  \\
$0\fdg79$ & 0.881  \\
$0\fdg89$ & 0.881  \\
$1\fdg20$  & 0.881  \\
\hline
\end{tabular}
\end{table}

\subsection{Establishing the antenna temperature scale} \label{gettingTa}

The preliminary antenna temperature calibration assumes that the noise
source has no frequency dependence over the range of our observations.
To check this we observed the standard radio continuum source 3C286 in
spectral line mode and measured the precise value of the receiver
noise source averaged in 50~\kms\ intervals around the Galactic 21-cm
line.  When the two linearly polarized channels are averaged the noise
source varies little with frequency, with a peak-to-peak fluctuation
about the mean of only $1.2\%$ over 600~\kms\ around the 21-cm
line. Thus the assumption of a constant calibration noise across our
band will not contribute a significant uncertainty to the \HI\
measurements when both polarizations are combined.

The absolute value of the noise source was checked using the flux
density standard 3C286 \citep{Ott1994} and the aperture efficiency
derived from the electromagnetic calculations (Sect.~\ref{calcside}).
From 13 measurements at two epochs we derive a calibration temperature
of $1.495\pm0.013$~K ($1\sigma$), a value that is in the ratio
$1.024\pm0.009$ to that measured in the laboratory. As 3C286 has
significant linear polarized emission this is for the average of the
two polarizations.  A smaller number of measurements on 3C295 gives a
result that is identical to within the uncertainties. We thus
increased the preliminary antenna temperatures by this ratio to place
spectra on an accurate $T_{\rm a}$ scale.  Note that this calibration,
derived from an external radio source, subsumes within it any
correction necessary for ohmic losses in the entire system.

\subsection{A check using the Moon} \label{mainBeamEff}

The $T_{\rm a}$ scale and value of $\eta_{\rm mb}$ were further
checked by continuum observations of the Moon, assumed to have a
constant brightness temperature of $225\pm5$ K at 1.4~GHz
\citep{Keihm1975} and taking into account that only a part of the
response pattern shown in Fig.~\ref{mainbeam} lies within the disk of
the Moon.  Measurements at several epochs give a measured to expected
ratio $0.97\pm0.025$ ($1\sigma$), where the uncertainty includes the
scatter in the measurements and in the assumed $T_{\rm b}$ of the
Moon.  Because we measure the Moon relative to the nearby sky, these
observations are not susceptible to radiation in the far sidelobes and
so are an uncomplicated test of both the $T_{\rm a}$ scale and
$\eta_{\rm mb}$ determination.  We conclude that our calibration does
not have systematic errors that exceed a few percent.

\section{GBT all-sky response} \label{sidelobes}

Calculating the stray 21-cm component requires knowledge of the
antenna response in all directions on the sky.  Important sidelobes
can occur at a level 50~dB below that of the main beam, and thus be
quite difficult to measure.  Nonetheless, they can contain several
percent of the telescope's total response because they cover a large
area on the sky.  In an analysis of the 21-cm beam pattern of the
Effelsberg 100-m antenna, \citet{Kalberla1980} found 70\% of the
response within 15\arcmin\ of the main beam, another 12\% in the range
$15\arcmin < \theta < 4\degr$, and the remaining 18\% at $4\degr <
\theta < 180\degr$.  An unblocked antenna like the GBT has much lower
sidelobe levels, but at 21~cm there is still an important contribution
to stray radiation due to spillover past the subreflector.  Note that
the sidelobe pattern for the GBT is not symmetrical about the main
beam, but for comparison more than 87\% of the response is within 15\arcmin\
of the main beam (Table~\ref{eta_mb_table}), with another 1.8\% in the range
$15\arcmin < \theta < 4\degr$.

To determine the all-sky response $P$ of the GBT we first used the
reflector antenna code (Sect.~\ref{calcside}) to find the calculated
response.  The results were used to set the aperture efficiency, main
beam efficiency, and the power pattern within $\theta < 1\degr$ of the
main beam.

These calculations also give a general picture of the location and
amplitude of the far sidelobes.  Key features of the sidelobes can
also be understood using simple near-field diffraction theory as
discussed in Appendix~\ref{secondary}.  Because the antenna code uses
only an approximation to the complex structure of the GBT, a more
accurate determination was accomplished by measuring the sidelobes
directly over much of the region within $60\degr$ of the main beam,
exploiting the Sun as a strong radio continuum source.  The Sun can be
treated as a point source relative to the angular size of most of the
structure in the sidelobes.

\subsection{Sidelobe measurements using the Sun} \label{sunscans}

\subsubsection{Observations}\label{obssun}

\begin{figure}
\centering
\includegraphics[angle=0,width=0.74\linewidth]{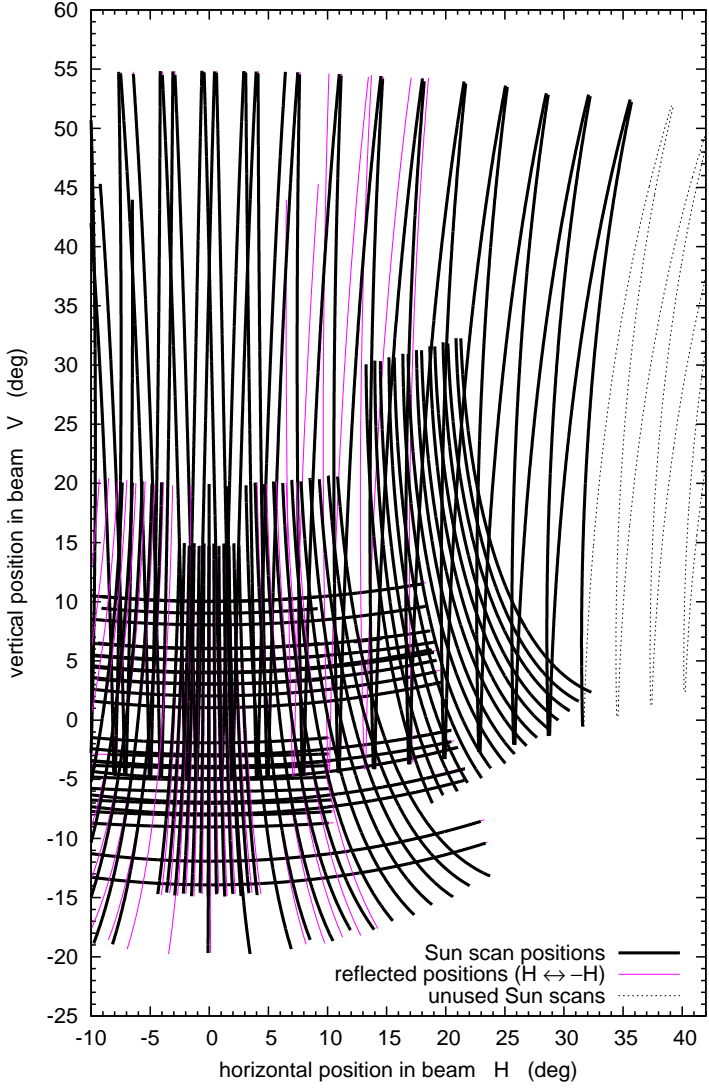}
\caption{Mapping the GBT sidelobe pattern using the Sun.  Heavy lines
  indicate the position of the Sun relative to the main beam during
  the scan.  Thin magenta lines indicate scans reflected in the beam's
  line of symmetry (i.e., with the $H$ coordinate replaced by $-H$).
  Dotted lines at high~$H$ show measurements that were not used
  because they showed no visible sidelobes above the noise.  The
  center of the subreflector is \subreflectordirection\ above the main
  beam.  }
\label{sunscan-pos-used-XY}
\end{figure}

The observations were made on three different occasions, during times
of the year when confusion with radio emission from the Galactic plane
would be minimized.  Data were taken over a 20~MHz band centered on
1420~MHz.  The mapping procedure consisted of raster scans, moving the
telescope either in elevation (``vertical'' scans) or azimuth
(``horizontal'' scans).
Figure~\ref{sunscan-pos-used-XY} shows the position of the Sun
relative to the main beam during the scans used to determine the
sidelobes, in terms of the vertical separation in elevation~$V$ from
the beam center, and the horizontal separation in the perpendicular
(cross-elevation) direction~$H$.  If the beam is pointed at the
horizon at an azimuth of $0^\circ$, then $V$ corresponds to elevation
and $H$~to azimuth.  We define $H = \theta \cos\phi$ and $V = -\theta
\sin\phi$, where $\theta$ is the angular distance from the beam center
and $\phi$ is the azimuthal angle around the beam, with $\phi =
0^\circ$ orthogonal to the antenna symmetry plane and $\phi =
90^\circ$ corresponding to the downward direction in elevation (away
from the GBT arm; see Fig.\ref{GBT_front} and Appendix~\ref{calcSri}).

The sunscans cover the most of the region $-20\degr < V < +55\degr$
and $-17\degr < H < +55\degr$.  The left-right reflection symmetry of
the telescope implies that a sunscan at~$-H$ should be identical to
one at~$H$.
The data were recorded at intervals of between 2 and~10~s,
corresponding to separations of a few arcminutes on the sky at the
adopted scan rates.

The initial data set consisted of ``vertical'' raster scans spaced
$2\degr-4\degr$ apart.  Among these, the ``distant vertical'' scans
lying beyond $H \sim 35\degr$ from the main beam (dotted lines at
right in Fig.~\ref{sunscan-pos-used-XY}) showed no evidence of
sidelobes above the noise level.  The theoretical values for these
scans are small enough to be consistent with a measured value of zero,
considering the uncertainties in the data.

The second data set consisted of ``horizontal'' scans spaced by about
$1\degr$ in~$V$.  The third set consisted of ``vertical'' scans, most
spaced by about $1\degr$ in~$H$, but including a group with
$30\arcmin$ spacing refining the coverage of the range $|H| < 3\degr$.
Although we were not able to measure the power pattern to large
negative angles~$H$, both the theoretical calculations and our
observations indicate that the pattern is symmetric about a vertical
line ($H = 0\degr$) through the center of the main beam.

\subsubsection{Data reduction and relative calibration of the
  observations of the Sun}\label{getfsun}

A baseline was set using the longest scans, assuming that the
sidelobes many tens of degrees from the main beam are relatively
negligible; the calculations suggest that away from the spillover lobe
the typical sidelobe has an amplitude of $-75$~dB with respect to the
main beam.  Backgrounds had to be subtracted, consisting of a large
constant component plus a smaller elevation-dependent component with a
slight amount of curvature which was well fitted by an exponential
which accounts for atmospheric emission variations with elevation.
The most distant vertical scans that showed no strong sidelobes were
used to examine the elevation dependence of the background signal in
the recorded data.  Both background components varied to some extent
from scan to scan, with the constant component also being slightly
different for the two polarizations.  Spikes in the data due to RFI
were also removed.
The numerous scan crossings were used to fix the amplitude of the
shorter scans.  Consistency of these scan crossings, and of the scans
symmetric about $H = 0\degr$, indicated an accuracy of approximately
10\% in the background-subtracted scans; the horizontal scans appeared
somewhat less accurate, possibly due to the variation in the height of
the horizon along a horizontal scan.

At 1.4~GHz the Sun can have significant temporal variations in its
emission.  Relative normalizations of the three datasets were
determined from the numerous scan crossing points.  We could not find
21~cm measurements of the solar flux for the periods of our
observations, and so we used the solar flux monitor measurements at
10.6~cm from the
DRAO\footnote{http://www.spaceweather.gc.ca/sx-11-eng.php} to check
that the relative normalization factors were consistent with the
ratios of the solar flux.  The second and third datasets were obtained
near the solar activity minimum, and required only a small relative
normalization factor; the first dataset had been obtained at at time
when the solar emission was approximately twice as large, requiring a
relative normalization factor of about two.

Scans were smoothed along the scan direction, and linear interpolation
was used to estimate values between the scans in a direction
approximately perpendicular to the scan direction.  Observations at
negative azimuth angles were ``flipped'' to positive azimuth,
providing additional coverage and consistency checks.  Where scans
crossed or approached closer than $30\arcmin$ to each other, a
weighted average was used at their positions to avoid any sudden
jumps.  Separate interpolations were made for each of the three
datasets, with a weighting approximately proportional to the density
of the observations on the sky.

The result is a map of the relative GBT beam pattern covering the
important sidelobes within the range $|H| < 32\degr$ (horizontal
extent) and $-20\degr < V < +55\degr$ (vertical extent).

\subsubsection{Determining the amplitude scale of the measured 
  sidelobes} \label{sunscanscale}

\begin{figure}
\centering
\includegraphics[angle=90,width=0.9\linewidth]{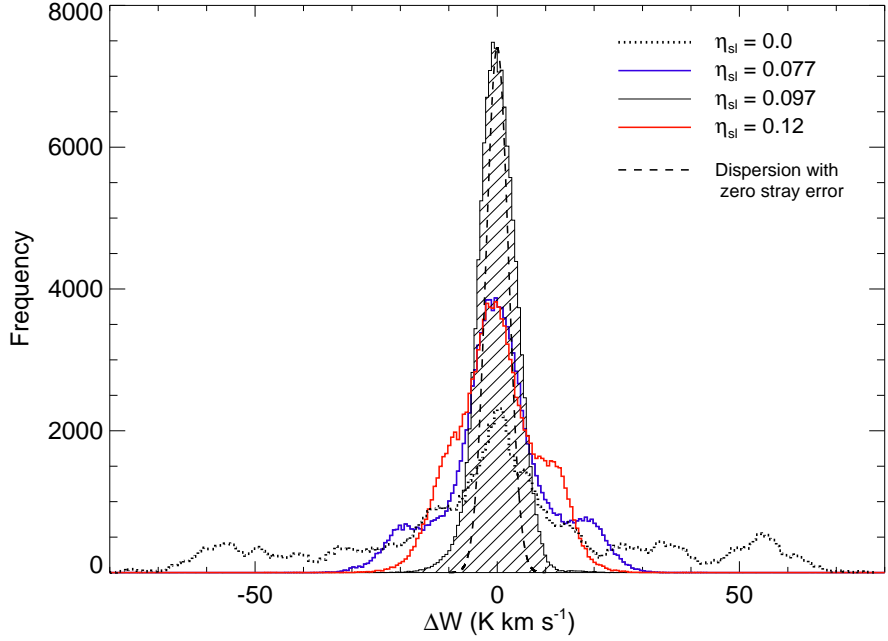}
\caption{Estimating the optimal value of $\eta_{\rm sl}$ using three
  repeated observations at more than $4\times10^4$ positions near the
  NEP, each repeat having different stray radiation.  For different
  assumed $\eta_{\rm sl}$, the difference between each observation of
  $W$ and the mean of the three was calculated for every
  position. Histograms of the resulting distributions of differences
  show the sensitivity to $\eta_{\rm sl}$, including the dotted curve
  for no stray radiation correction.  Correction is clearly
  beneficial, and a minimum dispersion occurs for $\eta_{\rm
    sl}\approx\slvalueNEP$.  Note that the dispersion does not quite
  reach the minimum predicted from line noise and baseline
  uncertainties alone (dashed line, see Sect.~\ref{specbase}). }
\label{nep_diff}
\end{figure}

The power pattern measured using the Sun needs to be scaled by an
amount determined below, because the absolute brightness of the Sun at
the times of observation is not known to the required precision.
Because the sidelobes are fixed with respect to the GBT, as a given
direction is observed the sidelobes will cross regions of different
21-cm $T_{\rm b}$ causing the stray radiation component to vary
throughout the day.  There are also different Doppler shifts.  We took
advantage of these temporal changes to estimate the scaling of the
measured sidelobes by mapping a large area around the North Ecliptic
Pole (NEP) containing $>4\times 10^4$ unique positions.  \HI\ spectra
were measured at each position on three separate occasions at
different ranges in azimuth and elevation.  For each position the
spectra were reduced, calibrated, and corrected for stray radiation
using Eq.~\ref{removestray} with a range of scalings which, when added
to the calculated part of the sidelobe pattern, gave a range of values
of $\eta_{\rm sl}$.  The amount of the telescope response in the
calculated portion of the far sidelobes, which is generally the region
at $\theta \gtrsim 60\degr$, is 0.0187.

A histogram of the differences between each of the three observations
of the integral over the \HI\ line profile ($W$, in K~\kms) and their
mean is shown in Fig.~\ref{nep_diff} for different representative
$\eta_{\rm sl}$.  The black dotted curve shows the large scatter in
the measurements when there is no correction for stray radiation.  The
stray radiation correction is quite effective: the dispersion
decreases to a minimum as the scaling increases, then subsequently
increases.  The optimum scaling of the empirical sidelobe pattern is
the one that minimizes the dispersion, near $\eta_{\rm sl} =
\slvalueNEP$.

A similar analysis, looking directly at differences in $T_{\rm mb}$
channel by channel instead of differences in the area $W$, was
performed on the NEP field and on other areas that were observed multiple times
during the course of the \citet{Martin2011} surveys.  For the spectra
in the NEP field, this analysis yields a value of $\eta_{\rm sl} = 0.1014
\pm 0.0018$, where the uncertainty is solely a measure of the
statistical accuracy of $\eta_{\rm sl}$; it does not take into account
the accuracy of the sidelobe pattern.  This places the above NEP $W$
result within 2$\sigma$, indicating the consistency of the two
analyses.  For all additional (non-NEP) spectra -- most of which only
have two repeat visits -- the $T_{\rm mb}$ analysis yields a value of
$\eta_{\rm sl} = 0.0947 \pm 0.0032$.  The statistical error was
obtained by a chi-squared analysis; to determine the number of degrees
of freedom, it was assumed that each continuously-observed ``chunk''
of sky (i.e., wherein all spectra have similar sky positions and LST
values) could be considered an independent measurement of~$\eta_{\rm
  sl}$.  There are proportionally a larger number of spectra in the
NEP field,
but they are all representative of a single patch of the sky with
three repeated observations, and statistically will have more similar
LST values --- observations of nearly the same patch of sky at nearly
the same LST do not yield entirely independent measurements
of~$\eta_{\rm sl}$.  Obtaining the best $\eta_{\rm sl}$ separately for
all nine subregions observed in the NEP survey yields values that are
all within 0.5$\sigma$ of the best overall NEP value, according to
their individual statistical errors, unlike the 13 separate non-NEP
regions, where the scatter is consistent with the statistical errors.
Thus the formal statistical error from the NEP measurement
underestimates its true error, which is probably closer to the non-NEP
error estimate, and so we decided that an average of the two values
(NEP and non-NEP) would yield a less biased result for~$\eta_{\rm
  sl}$.  Consequently, we have adopted $\eta_{\rm sl}= \slvalue \pm
0.0023$ to set the amplitude of the measured sidelobes on a correct
scale relative to the GBT main beam.

Note that $\eta_{\rm mb} + \eta_{\rm sl} \approx 0.98$; we cannot
account for about 2\% of the power.  Cases were tested where an
isotropic component $\eta_{\rm iso}$ was added to the sidelobes to
bring the total beam power closer to unity.  However, even $\eta_{\rm
  iso} = 0.01$ (i.e., an increase of 0.005 in the sidelobe power seen
by the sky) yielded significantly more overcorrection, with parts of
stray-corrected spectra going negative.
The formal uncertainty of 0.0023 refers to overall increases or
decreases in the measured sidelobe.  Modifications to the shape of the
measured sidelobes (e.g., an isotropic component; see also
Appendix~\ref{diffraction}), would yield a different ``best $\eta_{\rm
  sl}$'' value.  Considering possible variations yields a somewhat
larger error estimate of 0.005.

\subsection{Combining the calculated and measured power
  pattern} \label{combining}

The adopted power pattern $P$ is a combination of the calculations of
Sect.~\ref{calcside} with the measurements of the Sun of
Sect.~\ref{sunscans}.  Measured values at $\theta > 1\fdg2$ begin to
differ from the calculated values, while at $\theta < 0\fdg8$ the
coarse, half-degree spacing of the measurements is insufficient to
probe the smaller-scale beam structure there; for both these regions,
differences between measurement and calculation can exceed a factor
of~2.  However, for $\theta \approx 1\degr$, measured values ($P$
convolved with the Sun) typically agree to better than~20\% with
smoothed calculated values ($P$ convolved with a paraboloid to yield a
resolution of 0\fdg5) --- this is nearly the best that can be expected
from the inherent uncertainties in the measurements.  Therefore, for
$0\fdg8 < \theta < 1\fdg2$, we make a smooth switchover from the
(convolved) calculated $P$ to the measured one.  As we correct for
sidelobes only at $\theta \gtrsim 1\degr$, this switchover has a
negligible effect on the results.

The theoretical values for $P$ outside the region measured using the
Sun are small enough to be consistent with a measured value of zero,
considering the uncertainties in the data.  Some of the back sidelobes
are simply not accessible, being below the horizon; they are also not
used in our evaluation of the stray radiation
(Sect.~\ref{implementstray}).  Therefore at angles beyond where we
were able to probe directly, the theoretical calculations of $P$ were
again adopted with a smooth switchover at the outer 0\fdg5 edge of the
measured region to prevent any possible discontinuities in the
adopted~$P$.

\subsection{Properties of the adopted response pattern} \label{sunside}

\begin{figure}
\centering
\includegraphics[angle=0,width=0.74\linewidth]{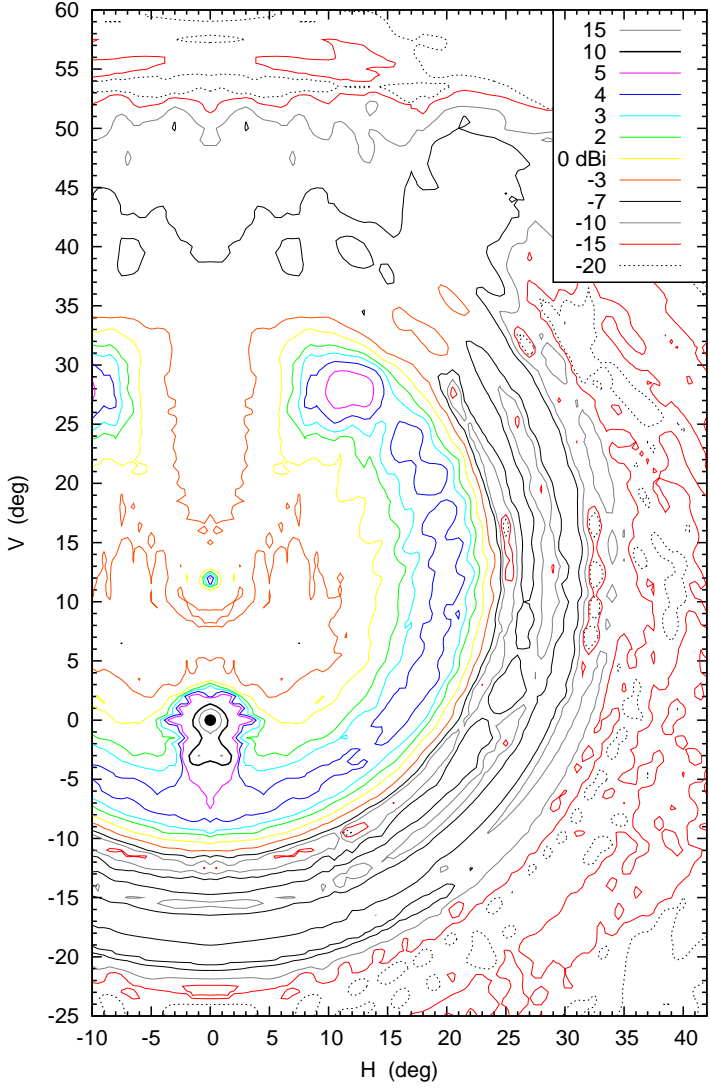}
\caption{Contours of the measured GBT far sidelobe gain~$G$, relative to
isotropic.
The main beam is at the origin (indicated by the solid dot), and would
peak at $+61.6$~dBi.  The $H$ and $V$ coordinates are the same as in
Fig.~\ref{sunscan-pos-used-XY}.  Elevation increases upward in the plot
and the direction to the center of the subreflector is at $V
=$\subreflectordirection.
The main ``spillover lobe'' is outlined by the yellow 0~dBi contour,
with the blue 4~dBi and magenta 5~dBi contours defining its ridge;
curving as a ring from below the main beam at $(H, V) \sim (0\degr,
-7\degr)$, it extends to about $(H, V) \sim (12\degr, 28\degr)$, beyond
which it is blocked by the screen on the telescope arm (note that this
blockage also yields a ridge stretching upward from the subreflector
center along the edge of the gap at an angle about $35\degr$ from the
vertical; this ridge passes through the end of the spillover lobe).
Complementary to this missing part of the spillover lobe caused by the
screen are the twin peaks below the main beam, at $(H, V) \sim
(\pm1\degr, -3\degr)$, visible in the heavy black 10~dBi and thin gray
15~dBi contours.
Outside the spillover lobe are three lower-amplitude rings (at radii
$\theta \sim 26\degr$, $29.5\degr$, and $33\degr$ from the subreflector
center).  
Inside is the Arago spot and surrounding rings centred at
\aragodirection\ on the axis above the main beam.
Some features are easier to see in the \hbox{3-D} plots of
Figs.~\ref{sunscan-sidelobes} and~\ref{sunscan-Arago-sidelobes}.
Outside the faint spillover rings and outside the region ($|H| <
35\degr, 20\degr < V < 53\degr$), the sidelobe levels are taken from
the antenna code calculations.  }
\label{mainbeam_and_sidelobe_contour}
\end{figure}

\begin{figure}
\centering
\includegraphics[angle=-90,width=1.0\linewidth]{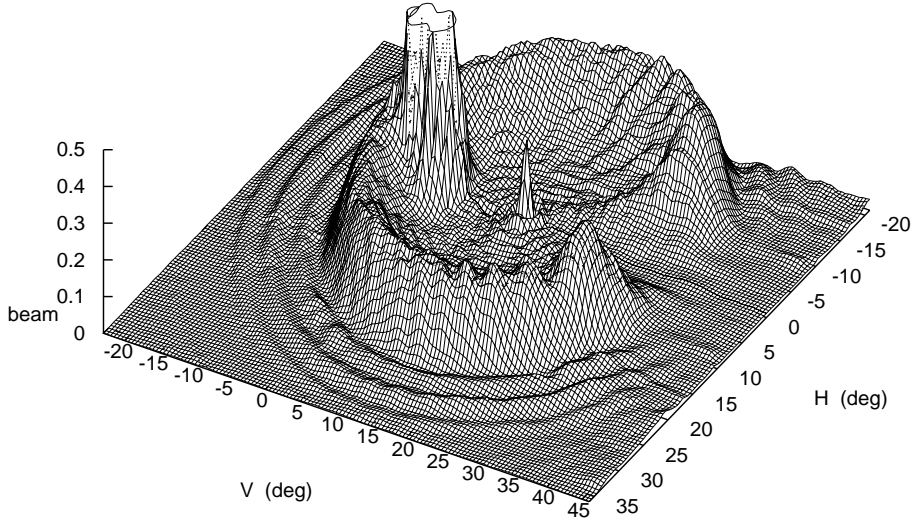}
\caption{GBT response pattern $P$, as measured from scans of the Sun,
  on a linear scale.  The $H$ and $V$ coordinates are the same as in
  Figs.~\ref{sunscan-pos-used-XY}
  and~\ref{mainbeam_and_sidelobe_contour}.  Note that this plot has
  been rotated so that its features are more easily visible --- the
  vertical coordinate~$V$ increases from left to right in this figure.
  The truncated peak at center-left in this figure comprises both the
  main beam at $(H,V) = (0,0)$ and the double-peak feature at $(H,V)
  \sim (\pm1\degr,-3\degr)$; the peak of the main beam would be {\it
    far\/} offscale at $P = 1.12 \times 10^5$.  Near the center of the
  spillover lobe, the Arago spot is visible.  The gap at high~$V$ in
  the spillover lobe and its surrounding rings arises from the
  presence of the reflecting screen at the subreflector edge.}
\label{sunscan-sidelobes}
\end{figure}

\begin{figure}
\centering
\includegraphics[angle=-90,width=1.2\linewidth]{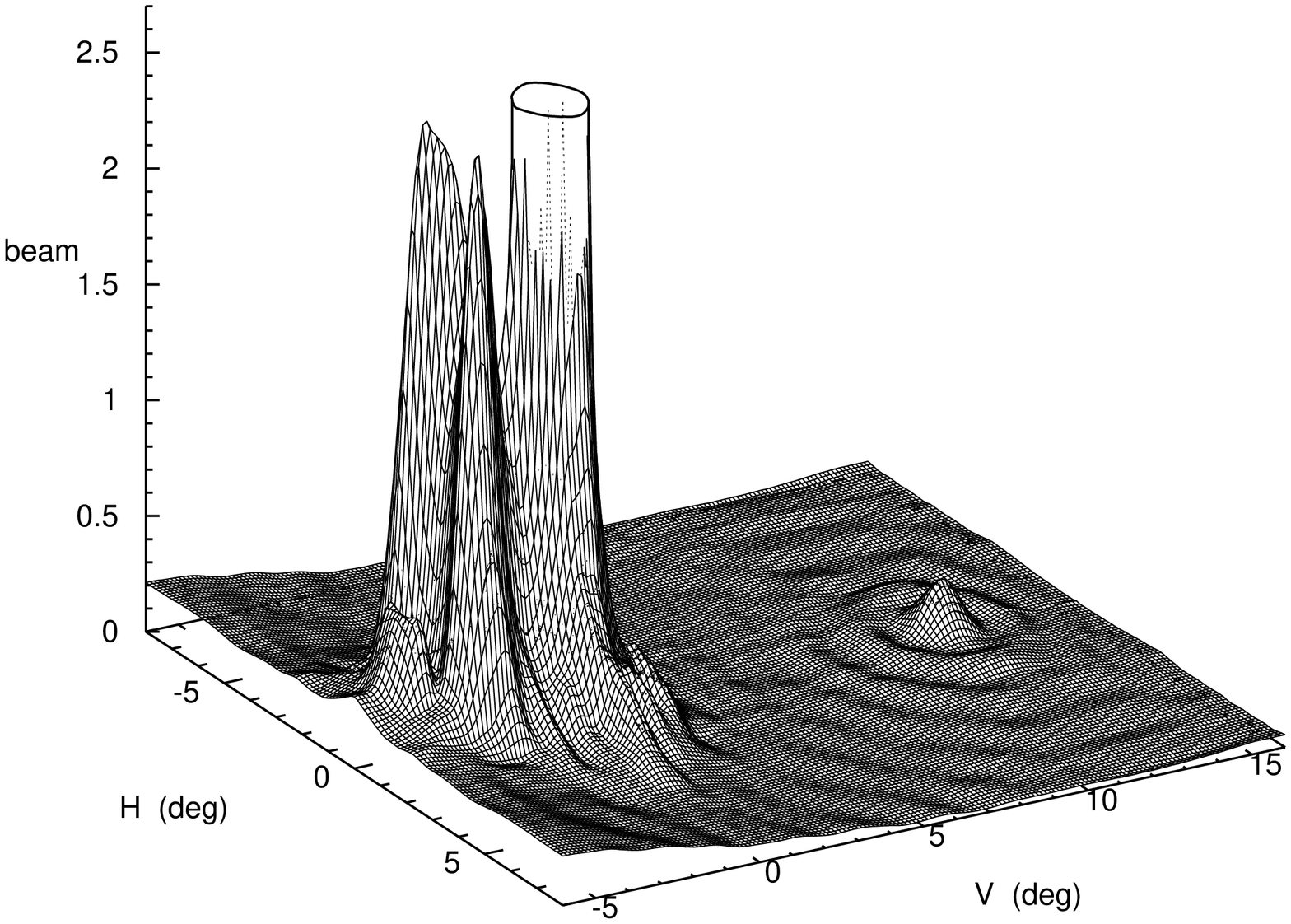}
\caption{Measured GBT response pattern $P$, as in
  Fig.~\ref{sunscan-sidelobes}, but for sidelobes closer to the main
  beam. The Arago spot, peaking at $\approx 0.25$, is visible at the
  right, along with its surrounding rings.  The double-peak feature at
  $(H,V) \sim (\pm1\degr,-3\degr)$ appears just to the left of the
  (truncated) main beam, with height $P \approx 2.6$. This contains
  the power scattered from the forward spillover lobe by the
  reflecting screen on the feed arm (Fig.~\ref{GBT_sub_screen}).  The
  peak of the main beam would be {\it far\/} offscale at $P = 1.12
  \times 10^5$. }
\label{sunscan-Arago-sidelobes}
\end{figure}

Figure \ref{mainbeam_and_sidelobe_contour} is a contour map of the
inner part of the derived GBT power pattern probed with the Sun; this
is $G$ in dBi.  The obvious symmetry about a vertical line through the
center of the main beam ($H=0\degr$) is by construction, as discussed
above in Sect.~\ref{obssun}.
Other displays of the GBT sidelobe pattern are given in
Figs.~\ref{sunscan-sidelobes} and~\ref{sunscan-Arago-sidelobes}; these
are on a linear scale and are for $P = 4\pi G$.

Away from the main beam, the power pattern is dominated by the forward
spillover sidelobe, the arc of radiation from the secondary feed
spilled past the subreflector (see Fig.~\ref{sunscan-sidelobes}).
In symmetric antennas such spillover sidelobes are symmetrical about
the main beam, but with the offset subreflector of the GBT
\citep{Norrod1996} it is centered roughly on the cone axis defining
the secondary, displaced by about $V = \spilloverdirection$ above the
main beam in elevation (it retains left-right symmetry but is not
circularly symmetric).
This spillover lobe results from near-field (Fresnel) diffraction of
the feed illumination from the sharp edge of the subreflector, which
can be thought of as a disk occulting the sky.  Our measurements of
the diameters of the main spillover lobe and the fainter rings outside
it are consistent with this, given the GBT geometry
(Appendix~\ref{secondary}).
The peak level along the ridge of this main spillover lobe is quite
low, about +5~dB above isotropic and therefore about $57$~dB below the
main beam, but because of the large area it contributes substantially
to $\eta_{\rm sl}$ and the stray radiation.

The gap in the spillover lobe at $(H,V) \sim (0\degr,30\degr)$ is
caused by the arm that supports the GBT subreflector, specifically a
reflecting screen attached to that arm (Fig.~3).  It deflects the
spillover radiation back into the main dish over a $40\degr$ segment
of the spillover lobe.  The reflected radiation emerges on the sky as
a pair of sidelobes well away from the main beam on the opposite side,
at about $H = \pm 1\degr, V = -3\degr$
(Fig.~\ref{mainbeam_and_sidelobe_contour}).  Each peak is elongated by
about $1\degr$, approximately along the azimuthal direction as seen at
the left in Fig.~\ref{sunscan-Arago-sidelobes}.  They have a peak
amplitude about ten times that of the spillover lobe, but comprise
only a small part of the total beam integral, comparable to the
portion of the major ring removed in the wedge, qualitatively
consistent with conservation of energy.  This is discussed in more
detail in Sect.~\ref{screen}.

A more minor, but interesting, feature in the GBT beam pattern is the
Poisson-Arago spot at $H =0\degr, V = $\aragodirection\ on the axis
above the main beam (Figs.\ref{mainbeam_and_sidelobe_contour} --
\ref{sunscan-Arago-sidelobes}).  As expected from simple near-field
diffraction theory, this is nearly the same direction as the center of
the subreflector (\subreflectordirection).  The width of the Arago
spot and the set of rings seen around it are also in detailed
agreement with simple diffraction theory (Appendix~\ref{aragospot}).

There is a feature in the (calculated) beam pattern located 
at $V \approx -96\degr$, with a peak amplitude about 57~dB 
below the main beam, slightly above isotropic 
(see Fig.~\ref{Sri_appendix_fig4}).  It
has a width of several degrees in $V$, is somewhat wider in $H$, and
is part of a ring, very asymmetric in both amplitude and position
about the main beam.  This sidelobe arises from spillover past the
edges of the main telescope reflector (Appendix~\ref{calcSri}).  Along
with this is another Arago spot in the direction from the prime focus
along the cone axis defining the primary \citep{Norrod1996}.  These
``backlobes'' contain roughly 2\% of the total power, i.e., roughly
20\% of the total sidelobes. However, because all but a tiny fraction
of these are always below the horizon, they do not ``see'' the sky and
so do not contribute significantly to the 21cm stray radiation.

\subsection{Implementation of the stray radiation correction}
  \label{implementstray}

Stray radiation was calculated for the GBT spectra following the
integral in Eq.~\ref{removestray}, with a program described more
extensively in Appendix~\ref{program}.  A model \HI\ sky was
constructed from the LAB survey data to give the input $T_{\rm
b}(\theta,\phi)$ for the integral, on a tiled grid in Galactic
latitude and longitude.  Given a specific GBT observation in a
particular direction at a specific time, the GBT response was
calculated for each tile location in the model sky more than $1\degr$
from the main beam.  The model sky spectrum was accumulated from all
directions above the local horizon after weighting by the beam
response, accounting for atmospheric attenuation, and making the
appropriate velocity shift.  This integrated stray radiation spectrum
was then subtracted from the observed $T_{\rm a}$ spectrum, before
correcting for atmospheric attenuation, and scaling by $\eta_{\rm
mb}$.
Implemented in the language C on a modern workstation, the program can
calculate the stray-radiation correction at a rate of several spectra
per second, or of order a square degree per minute for our mapped data
cubes.  The program is now available for use by any GBT observer.

\subsection{Examples of the effects of stray radiation}  \label{examples}

\begin{figure}
\centering
\includegraphics[angle=90,width=0.9\linewidth]{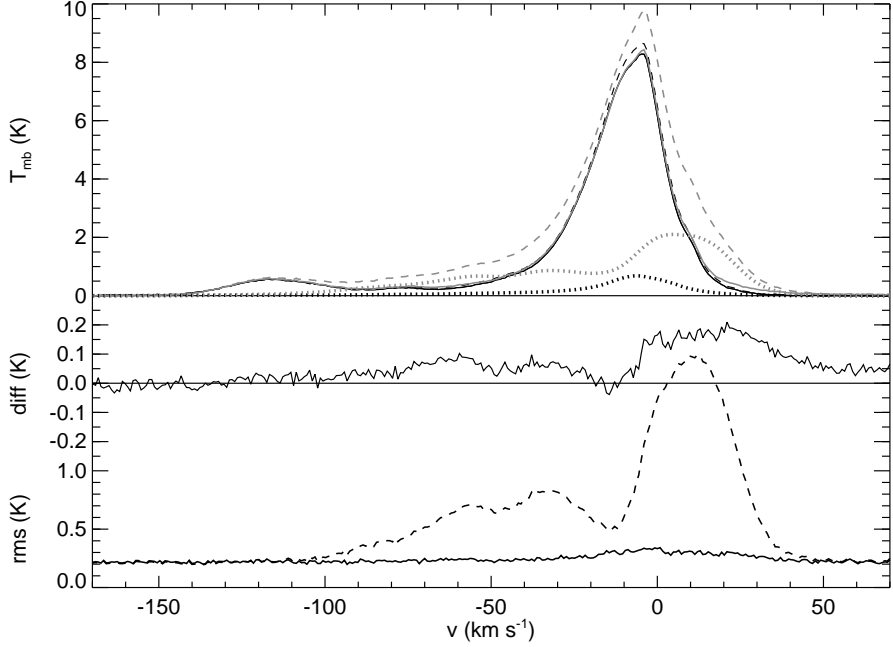}
\caption{
Example of correction for stray radiation for a subregion in the NEP field.
Upper panel: Dashed curves show spectra taken at different LST (black
and gray) with different stray radiation.  Each curve is the average of
280 contiguous \HI\ spectra on the $T_{\rm mb}$ scale; averaging is
essential for lowering the noise, to reveal more clearly the effects of
stray radiation.
Dotted curves: our calculation of the expected stray radiation.
Solid curves: spectra corrected for stray radiation, now well aligned.
Middle panel: mean difference of the corrected spectra (note that the
difference of the uncorrected spectra, or of the stray radiation, would
be far offscale).
Lower panel: spectra showing the rms of the 280 individual differences
before (dashed) and after (solid) the correction for the stray
radiation.  }
\label{fig_nep11_split4_stray}
\end{figure}

\begin{figure}
\centering
\includegraphics[angle=90,width=0.9\linewidth]{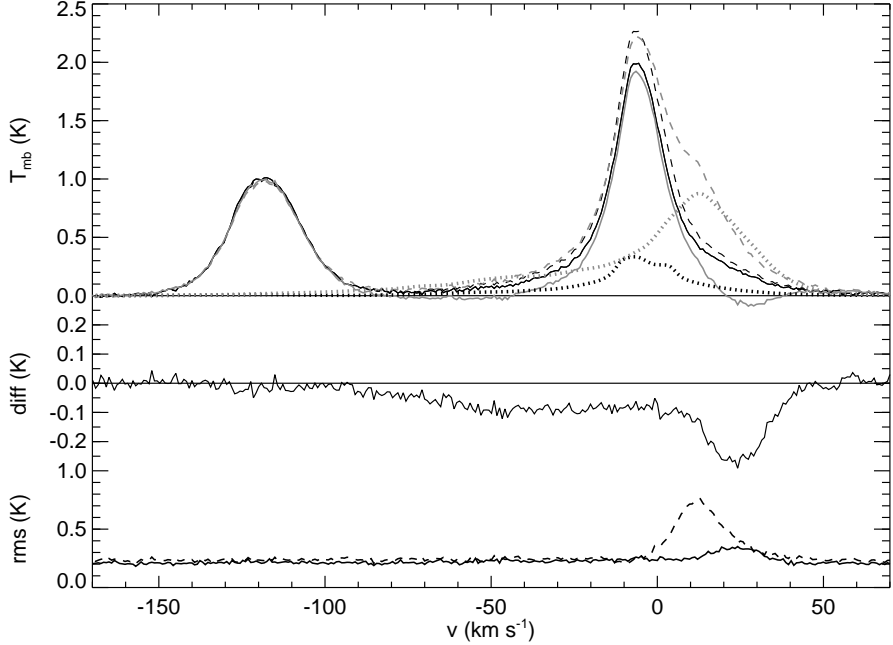}
\caption{ 
Like Fig.~\ref{fig_nep11_split4_stray}, but based on 264 contiguous
spectra in a subregion of N1.  The stray correction at one LST is
slightly too large, pushing the corrected spectrum slightly negative
in the high-velocity wing.  However, overall the correction is as
effective as in the NEP example; note that the middle and lower panels
are on the same scales as in Fig.~\ref{fig_nep11_split4_stray}.  }
\label{fig_N1_split3_stray}
\end{figure}

Examples of the stray radiation and our ability to remove it are shown
in Figs.~\ref{fig_nep11_split4_stray} and~\ref{fig_N1_split3_stray}.
Depending largely on where the spillover lobe is positioned on the
sky, the stray radiation correction ranges in peak amplitude from a
fraction of a K to several K.  Most stray radiation tends to lie
relatively close to zero velocity but is typically broader in velocity
than the corrected/intrinsic spectra; the total velocity range over
which stray radiation exceeds 0.1~K varies from a few tens of \kms\ to
about 150~\kms.

In these examples we compare spectra of the same region taken at
different LST so that they will have different stray radiation.  To
reduce the noise and reveal the subtle changes after correction for
stray radiation we have averaged many contiguous spectra that will
have rather similar stray radiation.
Figure~\ref{fig_nep11_split4_stray} illustrates a case where there are
large stray corrections that produce consistent results.  For one of
the epochs (gray lines in Fig.~\ref{fig_nep11_split4_stray}), the
stray correction removes excess emission on both wings of the main
peak, and the stray correction is significant all the way out to $v
\sim -120$~\kms.  Compared to the up to 2~K differences in the
uncorrected spectra, the difference between the corrected spectra is
an order of magnitude smaller (middle panel).  The difference after
correction is not, however, zero, and some residual effects of stray
radiation must still be present in the spectra.

Figure~\ref{fig_N1_split3_stray} provides another illustration from
the N1 region, a region of low signal where the stray correction can
have a {\emph{relatively}} large effect on the derived line profile
and its integral $W$.  Here, for one observation (gray lines in
Fig.~\ref{fig_N1_split3_stray}) the stray correction overcorrects
somewhat on either side of the main peak, yielding a slight dip below
zero near $v \sim 30$~\kms\ and $v \sim -60$~\kms.  However, the stray
correction still significantly reduces the difference between the two
observations and the amplitude of the rms difference spectrum, as
shown in the middle and lower panels of
Fig.~\ref{fig_N1_split3_stray}, respectively.  Note that these panels
are on the same scales as the corresponding panels in
Fig.~\ref{fig_nep11_split4_stray}; the residual effects remaining in
the corrected spectra are similar.
         
The quantification of the errors remaining because of imperfections in
the stray radiation correction, and the cumulative effects on $W$,
will be discussed further in Sect.~\ref{neptest} in the context of the
other sources of error.
Judging from many more comparisons of repeat measurements of different
fields, the largest {\it errors\/} in the stray radiation correction
tend to occur within a few tens of \kms\ of zero velocity, with a
typical amplitude of 0.1 to 0.2~K, although in the worst cases the
error can exceed 0.5~K.

\section{Error estimates}\label{testerrors}

In this section we evaluate the various contributions to the errors in
the GBT \HI\ spectra.  The effects of errors are manifested in the
non-reproducibility of measurements.  Comparisons between spectra can
be done on a channel by channel basis, $T(v)$, or for a line integral
$W$ (in K~\kms) of $T(v)$ over some velocity range.  Because of
different applications of \HI\ spectra, it is relevant to address the
errors for each metric.

\subsection{Line noise} \label{specnoise}

The individual survey spectra from \citet{Martin2011} have an rms
noise in emission-free channels of $\sigma_0 = 0.16$~K.  When the
spectra are interpolated into the data cube at Nyquist sampling this
is reduced to $\sigma_0 =0.11$~K \citep{Mangum07}.  The 21-cm line
emission itself can significantly increase the noise at velocities
where it is bright:
\begin{equation}
\label{specnoiseeqn}
\sigma(v) = \sigma_0 (1 + T(v)/T_{\mathrm{sys}}),
\end{equation}
where $T_{\rm sys}$ is the system temperature ($T_{\rm sys} \approx
20$~K, Sect.~\ref{GBTinstrumentation}).  For clarity, $T(v)$ is
referred to simply as $T$ for the remainder of this discussion.

All the GBT \HI\ observations measure spectra in two orthogonal linear
polarizations, labeled XX and YY, which should be receiving
essentially identical 21-cm emission and identical stray radiation.
Therefore, we are able to check the above equation by comparing
spectra in the two polarizations.  Both were processed in identical
parallel streams prior to removal of distinct instrumental baselines.
The difference in the baseline-subtracted XX and YY spectra provides a
good indication of the error inherent in a single measurement. It
eliminates the additional uncertainty arising from the stray radiation
subtraction which is in common, but includes channel noise and
baseline error.  This difference would also reveal any mis-calibration
of the two receiver channels and any real differences in the received
\HI\ signal because of differences in beam shape, pointing, or
sidelobes.  These latter effects, however, are thought to be small
compared to the other error terms.

\begin{figure}
\centering
\includegraphics[angle=90,width=0.9\linewidth]{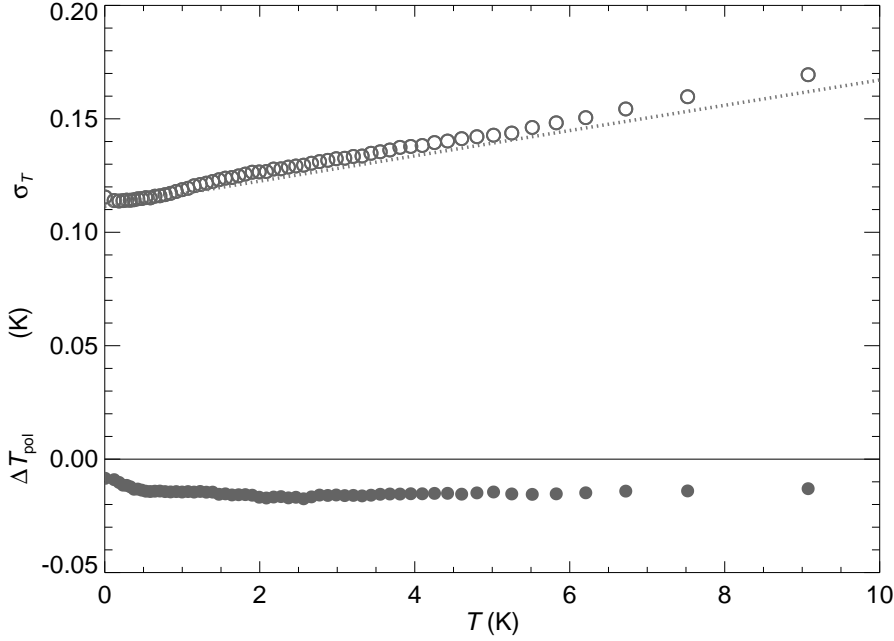}
\caption{Estimating the noise in a spectrum $T$ using $\Delta T_{\rm pol}$.
Open circles give the standard deviation of $\Delta T_{\rm pol}$ within bins
of $2\times10^5$ data points as a function of $T$, using spectra from
all NEP pointings.  This agrees closely with the prediction (dotted
line) from Eq.~\ref{specnoiseeqn} with $\sigma_{0}=0.111$~K and
$T_{\rm sys}=$20~K.
Also shown is the mean of $\Delta T_{\rm pol}$ for each bin.  }
\label{fig_nep_xx-yy_Tb}
\end{figure}

Because the errors in XX and YY are independent, and ultimately the XX
and YY spectra are averaged together to form $T = (T_{\rm XX}+T_{\rm
  YY})/2$, the estimator of interest in assessing errors in $T$ is the
dispersion $\sigma_T$, the standard deviation about the mean of
$\Delta T_{\rm pol} = (T_{\rm XX}-T_{\rm YY})/2$.
Figure~\ref{fig_nep_xx-yy_Tb} displays this dispersion based on all of
our NEP spectra.  The data over the $v$ range of $-50$~\kms to
$+25$~\kms have been binned by $T$, each bin containing $2\times10^5$
points.  The $T$-dependence of the standard deviation follows the
prediction from Eq.~\ref{specnoiseeqn}, overplotted in
Fig.~\ref{fig_nep_xx-yy_Tb} using
$\sigma_{0}=\sqrt{(\sigma_{0}^{\mathrm{XX}})^2 +
  (\sigma_{0}^{\mathrm{YY}})^2}/2 = 0.111$~K as measured in the
individual spectra and a typical $T_{\rm sys}$ of 20~K.

\subsection{Baselines}\label{specbase}

The mean of the distribution of $\Delta T_{\rm pol}$ within each bin,
also shown in Fig.~\ref{fig_nep_xx-yy_Tb}, is slightly offset from
zero, indicating a systematic difference in XX and YY that can be
attributed to imperfect baseline removal.  This offset -- though
well-characterized by a mean of $-15$~mK -- does vary from spectrum
to spectrum with a standard deviation of $\sim27$~mK.  This gives some
sense of the size of the baseline error in the polarization-averaged
spectrum $T$; it is so small compared to $\sigma(v)$ that it has
little effect on the dispersion described above.  However, because a
baseline error is systematic over many channels, it can accumulate as
a significant error in $W$.

\begin{figure}
\centering
\includegraphics[angle=90,width=0.9\linewidth]{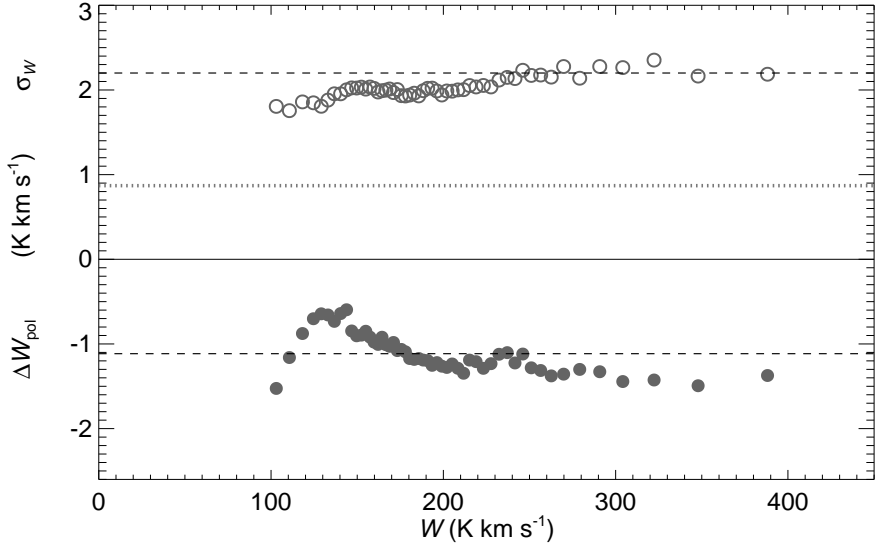}
\caption{Estimating the noise in the line integral $W$ using $\Delta
  W_{\rm pol}$.
Open circles give the standard deviation of $\Delta W_{\rm pol}$ within bins
of 2400 data points as function of $W$, using spectra from all NEP
pointings as in Fig.~\ref{fig_nep_xx-yy_Tb}.
This agrees closely with the prediction (dashed line) based on the accumulation of line noise (dotted line) and baseline errors. Note that the baseline error is dominant in measurements of 
 $W$, even though it is not
dominant for measurements of the spectrum $T$
(Fig.~\ref{fig_nep_xx-yy_Tb}).
Also shown are the data (filled circles) and the prediction (lower
dashed line) for the mean $\Delta W_{\rm pol}$.  }
\label{fig_nep_xx-yy_Wp}
\end{figure}

We illustrate this using the same NEP data, defining $W$ to be the
line integral over $N_{\rm ch} = 94$ channels of width $\Delta v =
0.8$~\kms\ over the range $-50$~\kms\ to +25~\kms.  We calculated the
dispersion $\sigma_W$ of $\Delta W_{\rm pol} = (W_{\rm XX} - W_{\rm
  YY})/2$ for bins in $W$ (2400 points each), plotting this in the
upper part of Fig.~\ref{fig_nep_xx-yy_Wp}.  The upper dashed line is a
prediction of this dispersion, combining in quadrature the minimal
channel noise $0.111 \sqrt{N_{\rm ch}} \Delta v$ (dotted line) and a
baseline error of $0.027 N_{\rm ch} \Delta v$ (both in K~\kms); note
that the latter is now the larger contribution, because of how it
accumulates systematically.
The larger values of $W$ often result from larger $T$ and thus larger
$\sigma(v)$, although this is not necessarily the case with very broad
lines.  The slight upward trend in the observed dispersion toward
larger $W$ could therefore be a result of the increasing contribution
of $\sigma(v)$ to the overall error.
For completeness, we show the mean of $\Delta W_{\rm pol}$ as well and
the prediction $-0.015 N_{\rm ch} \Delta v $~K~\kms.

The uncertainty in $W = (W_{\rm XX}+W_{\rm YY})/2$ should also be on
the order of 2~K~\kms; for optically thin emission, this corresponds
to an uncertainty of only $4 \times 10^{18}$~cm$^{-2 }$ in column
density.  Note, however, how this varies with the number of channels
used in the $W$ integral; it should be determined self-consistently
for the $W$ appropriate to each different region or application.

An independent estimate of the baseline errors can be made directly
from the third-order polynomial models used to fit the residual
baselines in the spectra (\ref{baselines}).  A Monte Carlo analysis
based on the uncertainties of the coefficients of the Legendre
polynomials yields errors in $W$ of $\sim0.7$~K~\kms\ over the same
$-50$~\kms\ to $+25$~\kms\ velocity range, suggesting that the
 errors derived from $\Delta W_{\rm pol}$ are an upper limit.
For the remainder of this paper we adopt this upper limit as it
includes potential systematic errors (e.g., offsets) between XX and YY
which are not detectable in their average, $W$.

The good agreement between the data and the predictions in
Figs.~\ref{fig_nep_xx-yy_Tb} and \ref{fig_nep_xx-yy_Wp} indicates that
we have a good understanding of the origins of the errors that arise
from noise and instrumental baselines and that there are not large
differences in the \HI\ signal measured in the two polarizations of a
single observation.
However, we still need to assess the errors arising from the stray
radiation correction (Sect.~\ref{neptest}), and any other time-varying
error contribution (Sect.~\ref{S6S8calib}).

\subsection{Stray radiation}\label{neptest}

As seen in the rms curves in Figs.~\ref{fig_nep11_split4_stray} and
\ref{fig_N1_split3_stray}, the baseline error and line noise are not
the entire story.  Changes that affect both XX and YY simultaneously
and systematically can only be diagnosed with repeated observations
at the same position.

The data from the NEP field, covering 44100 spatial pixels three times
each, are used here to examine the reproducibility of spectra with a
large range in $T$ and $W$.  This allows us to assess how the
uncertainty in the stray radiation correction contributes to the
overall error.  To estimate the errors in a single observation,
$T=(T_{\rm XX}+T_{\rm YY})/2$, for comparison with the results in the
subsections above, we examine the statistics of $\Delta T_{ij} = (T_i
- T_j)/\sqrt{2}$, where $i$ and $j$ denote two separate observations,
obtained at different LST and elevation as this region was mapped over
several months (2006/10 to 2007/01).

\begin{figure}
\centering
\includegraphics[angle=90,width=0.9\linewidth]{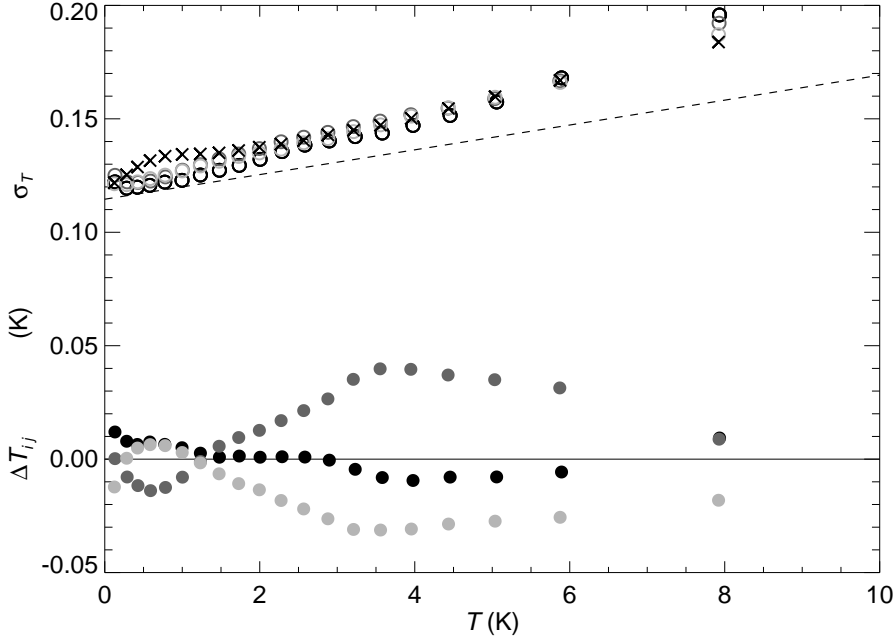}
\caption{Estimating the error in a spectrum $T$ following stray
  radiation removal, using repeated observations in the NEP field.
Open circles give the standard deviation of $\Delta T_{ij}$ within
bins of $2\times10^5$ data points as a function of $T$, using spectra
from the three pairs of NEP observations (dark gray: $3-2$, light
gray: $1-3$, black: $2-1$).
These lie slightly in excess of the prediction based on only line
noise plus baseline errors (dashed line).
Also shown is the mean of $\Delta T_{ij}$ for each bin for each $ij$
combination.
Both the excess and non-zero means arise from imperfect stray radiation
removal.
Crosses show a simple estimate of the total error which combines 6\%
of the rms spectrum of the stray radiation correction in quadrature
with the line noise.  }
\label{fig_nep_visits_Tb}
\end{figure}

The dispersion of $\Delta T_{ij}$ for data binned ($2\times10^5$ per
bin) in $T$, as in Sect.~\ref{specnoise}, is plotted as open circles
in the upper part of Fig.~\ref{fig_nep_visits_Tb} for each of the
three observation pairs.  The expected standard deviation from line
noise (plus a minimal baseline component of 0.027~K added in
quadrature) is plotted as a dashed line for $T_{\rm sys}=20$~K.  Also
shown as solid circles is the mean of $\Delta T_{ij}$ in each bin.  We
attribute both the excess of the observed dispersion above this
prediction and the non-zero means to errors or imperfections in the
stray radiation correction.  A simplistic estimate of the total error
can be obtained (crosses in Fig.~\ref{fig_nep_visits_Tb}) by including
some fraction (here 6\%) of the rms of the stray radiation
corrections, in quadrature with the line noise.

\begin{figure}
\centering
\includegraphics[angle=90,width=0.9\linewidth]{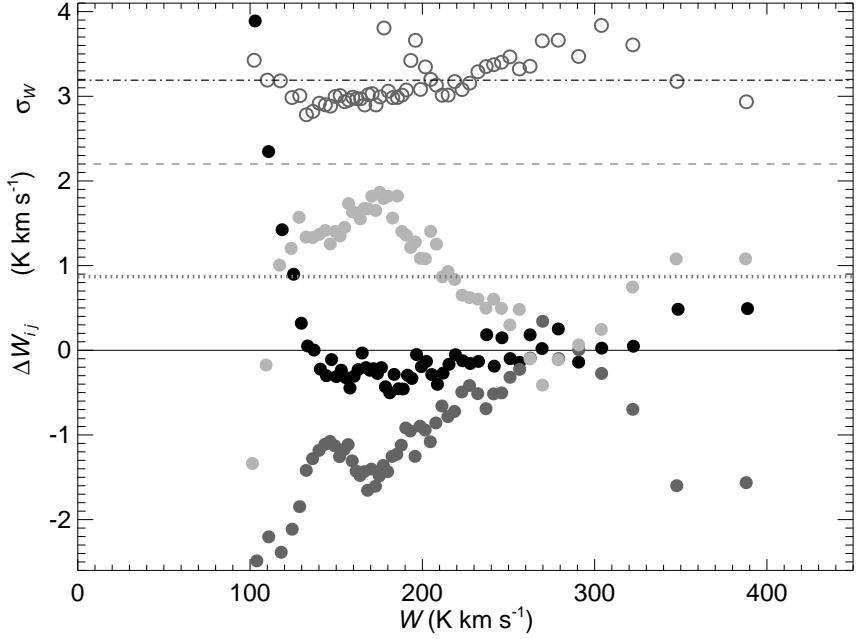}
\caption{Estimating the noise in the line integral $W$ using $\Delta
  W_{ij}$.
Open circles give the standard deviation of $\Delta W_{ij}$ over all
$ij$ combinations within bins of 800 data points as a function of $W$,
using repeated observations of spectra from all NEP pointings as in
Fig.~\ref{fig_nep_visits_Tb}.
This agrees closely with the prediction (dashed-dotted line) based on the
accumulation of systematic errors in the stray radiation and baseline
corrections added in quadrature with the line noise.
The line noise and baseline errors (dotted and dashed
lines from Fig.~\ref{fig_nep_xx-yy_Wp}) are shown for comparison.  Note that the error arising
from the stray radiation and baseline correction is dominant in measurements of
$W$, even though it is not dominant for measurements of the spectrum
$T$ (Fig.~\ref{fig_nep_visits_Tb}).
Also shown are the data (filled circles) for the means $\Delta W_{ij}$
for each $ij$ combination.  }
\label{fig_nep_visits_Wp}
\end{figure}

The mean difference does vary from spectrum to spectrum with a
standard deviation of
$\sim0.041$~K.  This is about 40\% of the typical dispersion in $T$
attributed to stray radiation.  This gives some sense of the size of
the {\it systematic} error from the stray radiation correction (see
also Figs.~\ref{fig_nep11_split4_stray} and
\ref{fig_N1_split3_stray}); it is small compared to the dispersion of
$\Delta T_{ij}$ but because it can be systematic over many channels,
it can accumulate as a significant error in $W$.  This is illustrated
using the same NEP data.
Again defining $W$ to be the line integral over $N_{ch} = 94$ channels
of width $\Delta v = 0.8$~\kms\ over the range $-50$~\kms\ to
+25~\kms, we calculated $\Delta W_{ij} = (W_{i} - W_{j})/\sqrt{2}$ for
bins (800 points) in $W$.  As shown in Fig.~\ref{fig_nep_visits_Wp},
the mean of $\Delta W_{ij}$ is typically $< 1.5$~K~\kms\ (lower points)
across all $W$ bins.
The dispersion in $\Delta W_{ij}$ over all $ij$ combinations is
plotted as open circles in the upper part of
Fig.~\ref{fig_nep_visits_Wp}.

This analysis of observations of $W$ in NEP, taken over a large range
of time and with very different stray radiation corrections, indicate
that the data are reproducible to an accuracy of 3~K~\kms\ in a single
mapping.  This represents an accuracy of 1\% to 3\%, depending on $W$.
As in Fig.~\ref{fig_nep_xx-yy_Wp}, the dotted line in
Fig.~\ref{fig_nep_visits_Wp} is the predicted dispersion from channel
noise alone, whereas the dashed line accounts for baseline errors as
well.  The upper dashed-dotted line is our prediction combining the
channel noise in quadrature with a systematic error from the stray
radiation correction of $0.041 N_{\rm ch} \Delta v$.  This systematic error
includes both baseline and stray radiation errors as it is not
possible to disentangle these here.
The prediction is in close agreement with
the observed dispersion.

The largest source of uncertainty in the stray radiation correction
appears to be from incomplete knowledge of the sidelobe pattern for
the telescope.  Other sources of uncertainty in the stray radiation
correction are discussed in Appendix~\ref{ssec:other}.
Note that the velocity range used in this example was selected to
accentuate errors from stray radiation.  Because there is little
significant stray contamination at $|v_{LSR}|>>0$~\kms, measurements
of $W$ for intermediate and high-velocity \HI\ components will have
uncertainties closer to those predicted from the appropriate $\Delta
W_{\rm pol}$.

For high-latitude regions of very low $T_{\rm b}$ and $W$, the stray
radiation from sidelobes overlapping \HI\ near the Galactic plane can
be as strong as the actual signal from the main beam at some
velocities.  The highest accuracy measurements can be obtained by
timing the observations so that the spillover lobe does not lie near
the bright \HI\ in the Galactic plane, since that minimizes the stray
radiation and the errors associated with its removal.  With our
knowledge of the extent and location of the sidelobes, we are able to
determine the ideal LST range and successfully apply this strategy to
the very faint ELAIS N1 field \citep{Martin2011}.
 
\subsection{Repeated observations of \HI\ calibration standards}
\label{S6S8calib}

The long-term reproducibility of the GBT spectra can be gauged using
our observations of standard \HI\ calibration regions, S6
(\citealp{Williams1973}, hereafter W73) and S8
(\citealp{Kalberla1982}, hereafter KMR), made over the period 2005/10
to 2008/03.  The integration time for each spectrum obtained was 180~s
and so these have much lower $T_{\rm sys}$ noise than the typical map
spectra.  Nevertheless they are affected by other sources of error,
principally in the stray radiation correction.  The spectra were
reduced using the standard procedures (Sect.~\ref{dataReduction}). For
each region we created an average spectrum $\left<T\right>$ from all
the data.

\begin{figure}
\centering
\includegraphics[angle=90,width=0.9\linewidth]{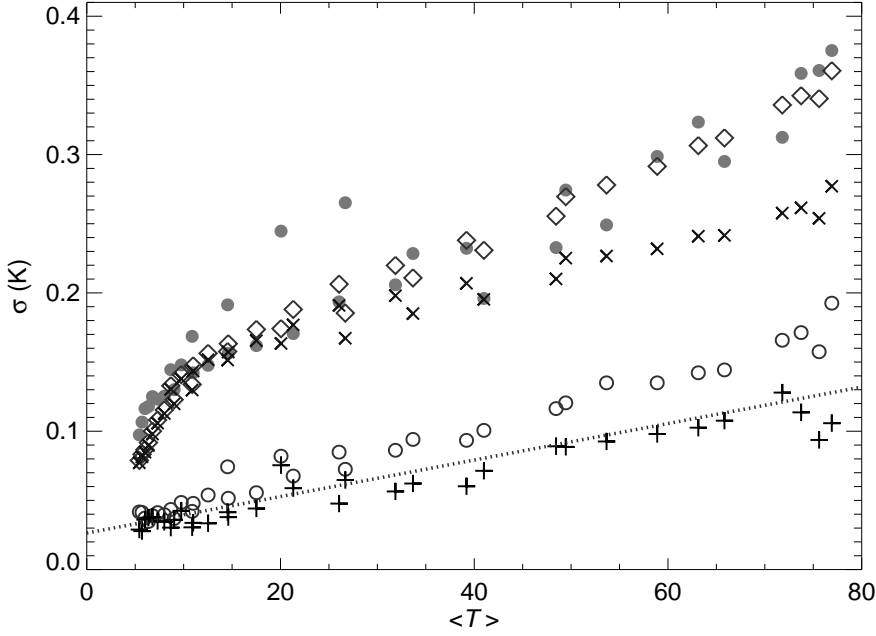}
\caption{
Estimating the noise in the spectrum $T$ of S8, using repeated
observations.
Plus symbols show the observed standard deviation of $\Delta T_{\rm 12}$
from channel by channel comparisons of two consecutive spectra, which
ought to have very similar stray radiation and baselines.  The
measurements agree closely with the prediction (dotted line) from
Eq.~\ref{specnoiseeqn} for line noise alone.
Open circles give the standard deviation of $\Delta T_{\rm pol}$; these lie
slightly above the prediction because of baseline errors.
The filled circles give the standard deviation of $\Delta T = T -
\left<T\right>$, which reflects all errors.
As a prediction of this, the crosses show the result of adding 7\% of
the rms spectrum of the stray radiation correction in quadrature to the
standard deviation of $\Delta T_{\rm pol}$.
The diamonds result from further addition of the effects of tiny 0.3\%
gain changes.  The S8 data are consistent with errors in stray
$\sim7\%$ and scaling (e.g., gain) errors of $< 0.3\%$.  }
\label{S8_repeats}
\end{figure}

There were 31 and 29 repeated observations of S6 and S8, respectively,
each consisting of two spectra, $T_{\rm 1}$ and $T_{\rm 2}$, from
consecutive 180~s integrations.  The stray radiation correction, and
also the baselines, ought to be very similar for the two spectra in
each pair.  We therefore examined $\Delta T_{\rm 12} = (T_{\rm 1} -
T_{\rm 2})/\sqrt{2}$ channel by channel.  The standard deviation of
$\Delta T_{\rm 12}$ for the 29 pairs in S8 is shown by the plus
symbols in the lower part of Fig.~\ref{S8_repeats}, for channels in
the velocity range used for the KMR calibration
(Sect.~\ref{S6S8calibAbs}).  As expected, this tracks closely the
prediction (dashed line) based on line noise alone from
Eq.~\ref{specnoiseeqn} with $\sigma_0 = 0.026$~K as measured in
emission-free channels, and $T_{\rm sys}=$20~K.

As a second check, we examined the standard deviation of $\Delta
T_{\rm pol}$ as in Fig.~\ref{fig_nep_xx-yy_Tb}
(Sect.~\ref{specnoise}).  These data are drawn as open circles in
Fig.~\ref{S8_repeats} and lie slightly above the dotted line because
of baseline errors, which for these long integrations with lower
$T_{\rm sys}$ noise have relatively more importance.

Next we examined the standard deviation of $\Delta T = T -
\left<T\right>$.  This is shown by the upper filled circles in
Fig.~\ref{S8_repeats}.  This somewhat larger dispersion (but note that
it is still less than 1\%) is from additional errors from the stray
radiation corrections and potentially tiny changes of gain with time.
We computed the rms spectrum of the actual stray radiation corrections
and take some percentage of this as a rough estimate of what the error
of the stray radiation correction might be.  The crosses in
Fig.~\ref{S8_repeats} show the result of taking just 7\% of the rms
spectrum and adding this error estimate in quadrature to the other
errors given by the standard deviation of $\Delta T_{\rm pol}$.
Although a simplistic description, it provides a reasonable
explanation for both the magnitude of the standard deviation of
$\Delta T$ and its dependence on $\left<T\right>$.

\begin{figure}
\centering
\includegraphics[angle=0,width=0.8\linewidth]{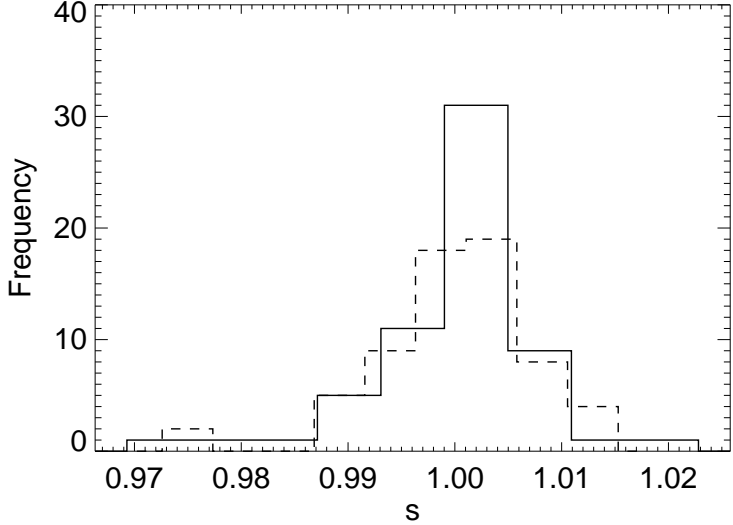}
\caption{The distribution of slopes $s$ from regressions of $T$ on
  $\left< T \right>$ for S6 (dashed) and S8 (solid) from observations
  made over 2.5 years.  The dispersions of 0.007 and 0.008,
  respectively, result from line noise, baseline errors, and errors
  in the stray radiation correction, and potentially from gain
  variations.}
\label{s6s8_T-T_histo}
\end{figure}

Pursuing another approach that might reveal, for example, gain
variations, we looked at the channel by channel correlation of $T$
with $\left<T\right>$, the slope of the regression being $s$.  We
carried out these regressions for data in the velocity ranges as
defined by W73 and KMR for their calibration.  By definition $s$ will
have a mean of unity, but it might vary from observation to
observation due to various errors.  Histograms of $s$ are presented in
Fig.~\ref{s6s8_T-T_histo}.  The standard deviations are 0.008 and
0.006 for S6 and S8, respectively.  The GBT spectra are thus
reproducible to better than 1\% and thus any changes in gain are
smaller than 1\% as well.  The smallest values of $s$ occurred for
observations taken during periods of rain, suggesting an additional
source of atmospheric opacity not currently included in our reduction
procedure; these outliers (one pair for S6 and two pairs for S8) were
omitted in the above analysis of $\Delta T$.

These two analyses of S8 are not decisive as to the presence of tiny
secular changes in gain along with the errors from the stray radiation
correction.  S6 provides an opportunity to discriminate because the
stray radiation is relatively weaker compared to the $T_{\rm mb}$
signal.  Repeating the same analysis as for Fig.~\ref{S8_repeats} for
the case of S6, we found that the dispersion in $\Delta T$, though
smaller than in Fig.~\ref{S8_repeats}, is too large and has the wrong
$\left<T\right>$ dependence to be explained by an error of $5-10\%$ of
the rms stray radiation correction, suggesting that gain changes of
order 0.5\% might be present.  In summary, a common explanation for S6
and S8 suggests gain changes of order 0.3\% to 0.5\% combined with
stray correction error of size 7\% of the rms stray radiation
correction.  The additional effect of 0.3\% gain changes in S8, added
in quadrature with the other errors, is illustrated by the diamonds in
Fig.~\ref{S8_repeats}.
We have also reproduced Fig.~\ref{s6s8_T-T_histo} for the case of S8
using simulated spectra starting with $\left<T\right>$ and adding line
noise, 7\% of the rms stray radiation correction, and 0.3\% gain
changes with independent random seeds, verifying the self-consistency
of this identification and quantification of the sources of error.
Note that in the NEP (analysis in Fig.~\ref{fig_nep_visits_Tb}) and
our other \HI\ survey regions the signal $T_{\rm mb}$ is so small that
any tiny gain errors would have a small effect.

This excellent reproducibility is to be contrasted to the assertions
by \citet{Robishaw2009} that the GBT spectra suffer from 10\% errors
in the calibrated gain.  Their analysis depended on solving
simultaneously for gain and several parameters describing the
amplitude of their adopted sidelobe pattern.  However, their sidelobe
pattern was only approximate, so that the resulting relatively large
errors in the stray radiation correction become confused as implying
significant changes in gain.  We have seen such an ambiguity above,
but at a level $<1\%$, and our joint analysis for S6 and S8 via
Fig.~\ref{S8_repeats} and Fig.~\ref{s6s8_T-T_histo} is consistent with
contributions from errors in the stray radiation correction and only
tiny secular gain changes.
The tiny gain changes found are consistent with what might be expected
from the effects of gravitational and thermal distortions on the passive
surface of the main paraboloid of the GBT, as described in
Sect.~\ref{GBTinstrumentation}.

\subsection{Error budget and benefits from repeated observations}\label{repeat}

The contributions to the total error in the \HI\ spectral data are
summarized in Table~\ref{error_summary_table} for both the $T$ spectra
and the integrated $W$.  In the case that these errors are all
independent, these can be added in quadrature to assess the total
error.

\begin{table*}
\caption{Summary of errors in GBT \HI\ spectral data}
\label{error_summary_table}
\centering
\begin{tabular}{lcc}
\hline \hline
Error type&$\sigma$\tablefootmark{a} of $T$&Error in $W$\tablefootmark{b}, accumulated\\
    &(K)&from $\sigma$ of $T$ (K \kms)\\
\hline
$T_{\rm sys}$ noise       & 0.25\tablefootmark{c}  & $\ldots$\\
line noise, $\sigma(v)$  & $\beta \sqrt{(4~{\mathrm s})/t}\ (1+T(v)/T_{\rm sys})$\tablefootmark{d} & $\sqrt{N_{\rm ch}}\ {\mathrm {rms}}(\sigma(v)) \Delta v$\\
baseline                 & $0.027$                 & $0.027 N_{\rm ch} \Delta v$\\
stray                    & $0.07 T_{\rm stray}(v)$ & $0.07 W_{\rm stray}$\\
scale\tablefootmark{e}   & $0.005 T(v)$           & $0.005 W$\\
\hline
\vspace{-6 pt}
\end{tabular}
\tablefoot{
\tablefoottext{a}{Except where otherwise noted, all $\sigma$ are on
the $T_{\rm mb}$ scale, adjusted for $\eta_{\rm mb} = 0.88$.}
\tablefoottext{b}{Error calculated for $N_{\rm ch}$ channels with
spacing $\Delta v$. Assuming that each $W$ error is independent, the
total $W$ error is the sum of these, in quadrature.}
\tablefoottext{c}{RMS noise in antenna temperature ($T_{\rm a}$) for 1
second of integration in a 1~\kms\ channel for two polarizations and
in-band frequency switching.}
\tablefoottext{d}{RMS noise of $T_{\rm mb}$ in a 0.8~\kms\ channel for
two polarizations. $\beta = 0.16$ for a single pointing and $\beta =
0.11$ for spectrum interpolated into the data cube.  Integration
times are $t=4$~s for our scanning strategy (see
Sect.~\ref{HIobsTechnique}). For long S6 and S8 integrations,
$t=180$~s (see Sect.~\ref{S6S8calib}).  $T_{\rm sys}$ is approximately
$20$~K.}  \tablefoottext{e}{Includes any gain variations.}  }
\end{table*}

Observations are often repeated to ``beat down the errors''. This is
certainly beneficial for the errors arising from line noise and even
for the baseline errors.  However, it is apparent that the dominant
source of error in $W$ for our NEP map is from imperfect stray
radiation removal.  Furthermore, it is not clear how the dispersion of
$\Delta W_{ij}$ is to be interpreted in assessing the errors in even a
single observation.  In the worst case, the measured dispersion could
be entirely due to the uncertainty in one of the observations, and so
the uncertainty in that single observation of $W$ would be $\sqrt{2}$
times larger, on the order of 4~K~\kms\ or up to a few percent of the
values typical in NEP.
An error of this size corresponds to roughly 10\% of the typical stray
radiation contamination for these NEP observations, compared to the
estimated 7\% in Table~\ref{error_summary_table}.  Note that
additional observations will not necessarily make the resulting
average more accurate.  One can hope that there will be some
cancellation, but it will depend on the accuracy of the stray
radiation correction for the times of observation.

\section{Absolute calibration} \label{absolutecalibration}

\subsection{Standard \HI\ calibration directions}
\label{S6S8calibAbs}

S8 at $(l,b) = (207\fdg00, -15\fdg00)$ is among the IAU primary \HI\
calibrators and has been studied extensively by KMR using the
Effelsberg 100-m telescope.  To compare our repeated GBT observations
with KMR we computed the line integral $W$ over the prescribed
velocity range $-5.1 < v < +22.3$~\kms, finding $W_{\rm GBT} =
831\pm5$~K~\kms, the standard deviation being consistent with the
expectation from Fig.~\ref{S8_repeats}.  At the same angular
resolution, KMR found $W_{\rm Effel} = 846\pm14$~K~\kms, the estimated
error combining a general scale uncertainty of 1.5\% and an
approximate 0.2~K systematic uncertainty from their stray radiation
correction.  Thus $W_{\rm Effel}/W_{\rm GBT} = 1.018\pm0.018$.  We
conclude that our independent calibration procedure produces 21-cm
spectra in agreement with this detailed previous calibration at the
level of the uncertainties, 2\%.

S8 is a calibrator (along with the secondary calibrator S7) for the
Leiden/Dwingeloo Survey which forms the northern part of the LAB
survey.  The LAB resolution after regridding the spatially-sampled
data is about $40\arcmin$ \citep{Kalberla2005}.  To study changes in
the spectrum with angular resolution, KMR made a small $1\degr$ map
sampled on a rectangular grid at $5\arcmin$ intervals centred on S8.
Similarly, we made a $1\fdg5$ map with our standard scanning
setup. Errors from the stray radiation correction and baselines ought
to be fairly uniform over the map.  Convolving our map to ever lower
resolution, we reproduced the resolution dependence found by KMR
(their Fig.~5).  From the GBT resolution to the LAB resolution, $W$
increases by a factor $1.017\pm0.009$. Using this to scale our pointed
observations gives $W = 845\pm9$~K~\kms, to be compared to
$849\pm9$~K~\kms\ independently from our convolved small map, and
$856\pm14$~K~\kms\ found by KMR.  The value computed from the LAB cube
(with the Wakker correction -- see the following section) is
$844\pm12$~K~\kms, in close agreement at less than the 1\% level.

The other region that we measured repeatedly was S6 at $(l,b) =
(1\fdg91, 41\fdg42)$.  Determining $W$ over the range $-6.86 < v <
+5.86$~\kms\ (approximating W73; see KMR for a discussion) gives
$W_{\rm GBT} = 289\pm2$~K~\kms.  From a $1\fdg5$ scanned map we find
that the correction to the $35\arcmin$ ($40\arcmin$) angular
resolution of Hat Creek (LAB) is $0.990\pm0.006$ ($0.993\pm0.006$) giving
$286\pm3$~K~\kms ($287\pm3$).  The value directly from the convolved map is
$289\pm3$~K~\kms ($290\pm3$).  S6 is not a primary standard and has not been
measured as accurately as S8.  W73 reports $299\pm22$~K~\kms which is
probably systematically several percent high because it has not been
corrected for stray radiation.  Within the large 7\% uncertainty, it
agrees with our measurement.  The value computed from the LAB data,
interpolated to the S6 position, is $292\pm5$~K~\kms, again in close
agreement with our measurements at the 2\% level.

\subsection{Comparison with the LAB survey}\label{LABscale}

For comparison with the LAB Survey, the GBT \HI\ maps were convolved
to the $40\arcmin$ angular resolution of that survey.  The convolution
results in spectra with a negligible error from line noise, but any
errors from the baseline and stray radiation corrections are not
significantly reduced.  For repeated observations of a region, we used
the average spectrum.

Where the GBT observations completely cover a LAB beam, we
interpolated the convolved GBT data to the LAB positions in Galactic
coordinates ($0\fdg5\times0\fdg5$ grid in $l$ and~$b$) and to the
slightly coarser velocity grid.
The LAB and convolved-GBT spectra almost always exhibited very similar
spectral features, usually with a small difference in overall scale.
The line integral $W$ is used to quantify this scale; $W$ is defined
as the integral over all channels with $T > 1$~K in either spectrum.

We performed a regression of $W_{\rm LAB}$ on $W_{\rm GBT}$, and since
the LAB-survey-cube points at high Galactic latitudes~$b$ are not
independent, weights of $\cos\,b$ were applied.  This regression
yields
\begin{equation}
W_{\rm LAB} = ( 1.0298 \pm 0.0023 ) \times W_{\rm GBT} + 
( 1.96 \pm 0.21 ) \, {\rm K} \, \kms.
\label{lab_match_with_a}
\end{equation}
\citet{2011Wakker} has recommended that a Gaussian of peak amplitude
$0.048$~K, FWHM${} = 167$~\kms (equivalent to 8.5 {\rm K} \kms\ over
the entire LAB velocity range), and center $v = -22$~\kms\ be subtracted
from {\it all\/} LAB spectra.  After making this ``Wakker correction''
to the LAB spectra we find
\begin{equation}
W_{\rm LAB} = ( 1.0248 \pm 0.0022 ) \times W_{\rm GBT}
+ ( 0.71 \pm 0.20 ) \, {\rm K} \, \kms.
\label{lab_match_with_aWC}
\end{equation}
The intercept is now significantly lower, and so our data support the
application of the Wakker correction to the LAB Survey data.  The
following analysis adopts this correction.  However, note that this
correction is statistically based and might be different in different
regions of the survey and from spectrum to spectrum.

If we adopt the hypothesis that the true intercept is zero, then
\begin{equation}
W_{\rm LAB} = ( \labvsgbtratio \pm 0.0012 ) \times W_{\rm GBT}.
\label{lab_match_zero_a}
\end{equation}
We conclude that overall the LAB scale is about 3\% higher than for
the GBT.  This is taken into account in the model \HI\ sky used for
stray radiation correction; see Appendix~\ref{updated-program}.

\begin{figure}
\centering
\includegraphics[angle=-90,width=1.\linewidth]{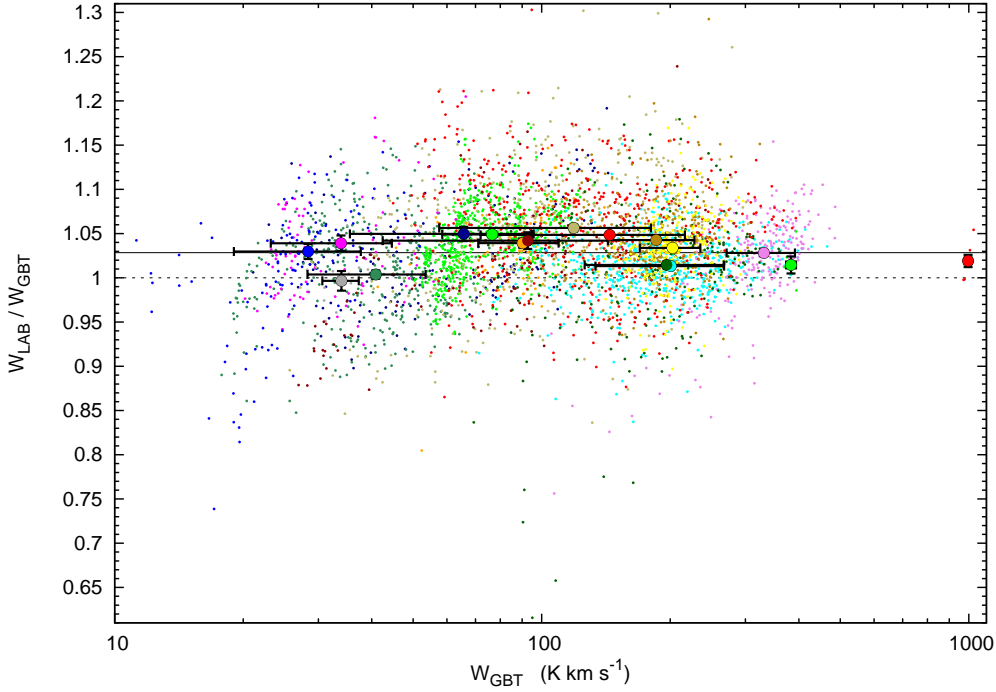}
\caption{ $W_{\rm LAB}/W_{\rm GBT}$ for every spectrum that could be
  compared (dots, color-coded by region).  Filled circles show, for
  each of the observed regions, the weighted average ratio and its
  uncertainty (smaller than the filled circle for the typical uncertainty
  of ${\sim 0.004}$ in the regional averages), along
  with the region's average $W_{\rm GBT}$ and its dispersion (indicated 
  by the horizontal errorbars).  These
  comparisons suggest that a single scale factor (horizontal solid line,
  from Eq.~\ref{lab_match_zero_a}) might be too simplistic.  }
\label{lab_ratio}
\end{figure}

A single scale factor or a single intercept might be too
simplistic.  This is illustrated in Fig.~\ref{lab_ratio} which shows
the ratio $W_{\rm LAB}/W_{\rm GBT}$ for every spectrum compared.
In agreement with the conclusions of \citet{Higgs2005}, we find that
the LAB Survey appears to have random ``calibration'' errors,
typically of a few percent.  About 20\% of the LAB spectra appear to
be mis-scaled (up and down) by between 5\% and 10\% and another 7\% of
the spectra by more than 10\%.  There are a few very large outliers
where a LAB spectrum has been mis-scaled by as much as 30\%; these
outliers were not included in the above analysis (though they
{\em are\/} plotted in Fig.~\ref{lab_ratio}).  Errors in the LAB
Survey are discussed in more detail in Appendix~\ref{LAB_errors}
(note that {\em random\/} LAB calibration errors tend to cancel out 
in the stray radiation calculation).

Also shown in Fig.~\ref{lab_ratio} are the weighted average ratios and
their uncertainties for the 17 regions that we mapped, as
obtained from regressions such as in Eq.~\ref{lab_match_zero_a} for
each region. 
The differences from region to region are significantly more than the
formal errors, suggesting that there might be systematic normalization
errors of a few percent in the LAB survey data which vary with
position in the sky.  
However, we have not found any systematic {\em trend\/} in the 
$W_{\rm LAB}/W_{\rm GBT}$ ratio as a function of position.

Fig.~\ref{lab_ratio} does not show any clear trend of 
$W_{\rm LAB}/W_{\rm GBT}$ or its dispersion vs.~$W$.  The S6 and S8 maps 
discussed above (Sect.~\ref{S6S8calibAbs}) have the largest $W$ values; 
based on six independent comparison positions in each of them, we
find $W_{\rm LAB}/W_{\rm GBT} = 1.015 \pm 0.010$ and $1.019 \pm 0.007$
for S6 and S8, respectively.  Restricting the comparison to the four
positions farthest from the GBT map edges reduces the ratios to $1.007 \pm
0.008$ and $1.010 \pm 0.007$, respectively.  Though smaller than the
overall ratio, these are still compatible with a 3\% difference
between LAB and GBT scales.

\subsection{Comparison with Lyman-$\alpha$ measurements of \HI} \label{Lyalpha}

Using a sample of 28 bright QSOs and AGN as targets,
\citet{2011Wakker} have compared measurements of the foreground
Galactic \nh\ as determined by UV spectroscopy in the Lyman-$\alpha$
absorption line with measurements of \nh\ from 21-cm GBT observations
in the same directions.  The GBT data were reduced as described in
this paper, and the correction for stray radiation was often
significant.  Third-order polynomials were fit to channels between
$-300 \leq v_{LSR} \leq -150$~\kms\ and $+100 \leq v_{LSR} \leq
+200$~\kms. For some directions showing emission from high-velocity
clouds the lower velocity range was changed to $-300 \leq v_{LSR} \leq
-200$~\kms.  Each 21-cm spectrum was examined for quality and a
third-order polynomial baseline was found to be a good fit to the
instrumental baseline.  The ratio \nh(Ly$\alpha$)/\nh(21~cm) $ =
1.00\pm0.07$ (1$\sigma$) indicating excellent agreement between the
two entirely independent sets of measurements and independent tracers
of \HI.  Some of the scatter must certainly result from structure in
\HI\ within the GBT beam as there is an enormous difference between
the angular scales sampled by the UV absorption and 21-cm emission.
We take this result as confirmation of the accuracy of our overall
calibration procedure and an indication that the total error in GBT
values of \nh\ must be significantly less than 7\%.

\section{Summary and conclusions}  \label{conclusions}

This paper describes the results of a program to develop a calibration
procedure that allows accurate measurement of 21-cm \HI\ spectra with
the Green Bank Telescope of the NRAO.  Using a combination of
measurement and calculation we have developed a model for the all-sky
response of the telescope and use it to correct for stray radiation in
the GBT 21-cm spectra.

Several methods were used to estimate errors in the final spectra.
Stochastic noise and instrumental baseline uncertainties are well
understood and can be made quite low in long integrations (Sect.
\ref{specnoise}).  The overall calibration to an absolute brightness
scale appears to be correct to within a few percent as judged against
measurements of standard radio continuum sources and the moon
(Sect.~\ref{aperturebeamefficiency}).

The correction for stray radiation is the most uncertain, and can
produce errors at any given velocity of as much as 0.5 K, though these
occur mainly at $|v_{LSR}|\lesssim 20$ \kms\ (Sect.~\ref{examples}).
Errors in the total $W$ caused by errors in the stray radiation
correction are typically less than 3~K~km~s$^{-1}$ or 4~K~\kms\ in the
worst case (Sects.~\ref{neptest} and \ref{repeat}).  These are
equivalent to an error in optically thin \nh\ of $\sim5 \times
10^{18}$ and $7 \times 10^{18}$ cm$^{-2}$.
Overall the system is quite stable.  We see no evidence for the ``10\%
gain fluctuations'' reported by \citet{Robishaw2009}.  An independent
investigation of the \HI\ content of nearby galaxies with the GBT
\citep{Hogg2007} achieved accuracies of $3.5\%$, where the main source
of error was baseline uncertainties.

The corrected GBT data are in good agreement with other measurements
of \HI. Most interestingly, the GBT data give, on average, identical
values of \nh\ as those derived from Lyman-$\alpha$ measurements
towards a sample of AGN and QSOs \citep[][and our
Sect.~\ref{Lyalpha}]{2011Wakker}.  The GBT data also agree to within a
few percent with previous measurements of ``standard'' \HI\
calibration directions and with other \HI\ observations at lower
angular resolution.  We have shown, however, that applying the
``Wakker correction'' to the LAB survey \citep{2011Wakker} improves
the agreement between that survey and the GBT spectra.

For the GBT at 1.42 GHz we find that 0.88 of the telescope's response
is within $1\degr$ of the main beam, with most concentrated within
$0\fdg2$ of the main beam, while another $0.098\pm0.005$ of the
telescope's response is in sidelobes more than one degree from the
main beam.  We thus account for $0.978 \pm 0.005$ of the telescope's
response including that which always lies on the ground, although the
{\em uncertainty\/} of $\pm0.005$ applies only to that part of the
sidelobe that sees the sky: the part that sees only the ground cannot
be measured, only calculated.  These results are very constrained.
Given the measurements of the sidelobes there are only two parameters
that can be varied: $\eta_{\rm mb}$ and $\eta_{\rm sl}$.  The value of
$\eta_{\rm sl}$ has been optimized from the observations
(Sect.~\ref{sunscanscale}); any change in $\eta_{\rm mb}$ would
increase discrepancies with other measurements
(Sect.~\ref{testerrors}).  Attempts to add an isotropic component to
the GBT beam pattern gave unacceptable results for the stray
correction, implying $\eta_{\rm iso} < 0.01$.  Strictly speaking, only
the part of the isotropic component lying above the horizon is
constrained by this analysis.  However, the magnitude of any extra
component of P in a backlobe is also strongly constrained by the low
receiver $T_{\rm sys}$.  Given that our model already has nearly $2\%$
of P intersecting the ground, any additional component that lies on
the ground must have much less than $1\%$ of the total P.

The calibration techniques are illustrated using data from \HI\ maps
made in connection with studies of interstellar dust and the cosmic
infrared background \citep{Martin2011, Abergel2011}.  Overall, the
data reduction process described here reduces systematic uncertainties
in GBT \HI\ spectra by at least an order of magnitude. Extra precision
might be obtained with a better model for the far sidelobes including
the possibility of reflections from the ground, allowance for the
slight changes in telescope geometry with elevation angle, and more
detailed consideration of atmospheric opacity during periods of rain.
Given the unblocked optics of the GBT it should be possible, in
principle, to construct a receiver for 21-cm work with negligible
forward sidelobes and a stray radiation component that is likewise
negligible.  This could be achieved in the near future with
phased-array feed receivers \citep{2010landon, 2010jeffs}.

\begin{acknowledgements}

We thank Ron Maddalena, Rick Fisher, and Roger Norrod for many helpful
discussions and Ron Maddalena for a critical reading of the manuscript.
Bob Anderson supplied Fig.~\ref{GBT_sub_screen}.
The National Radio Astronomy Observatory is a facility of the National
Science Foundation, operated under a cooperative agreement by
Associated Universities, Inc.
This work was supported in part by the Natural Sciences and
Engineering Research Council of Canada.

\end{acknowledgements}

\bibliography{stray_bib.bib}

\begin{appendix}

\section{Calculation of the GBT antenna pattern} \label{calcSri}

The main beam and the sidelobes of the GBT at 1.40 GHz were calculated
using a reflector antenna code developed at the Ohio State University.
This code, called the Antenna Workbench \citep{1974Kouyoumjian,
  1979Lee, 1985Lee, 1990Lee}, can be used to analyze single or
multiple reflectors.  It is capable of calculating both near-field and
far-field radiation patterns.  The theoretical approach is based on a
combination of Geometrical Theory of Diffraction (GTD) and Aperture
Integration (AI) techniques.  Typically, AI is used for computing the
main beam and the near sidelobes while GTD is used for computing the
far sidelobes including the backlobes.  For near-field calculations,
GTD is used in some cases for the whole region including the near axis
region.  In addition, the code has the conventional Physical Optics
(PO) method option, where the currents on the reflector surface over a
two-dimensional grid system are integrated to obtain the radiation
pattern.  The code usually uses the PO option for antennas with
multiple reflectors.  The Antenna Workbench was modified specifically
for the large main reflector of the GBT.  In the present analysis, the
multiple reflector option using the PO technique was not used because
it does not include edge effects and hence cannot compute the far
sidelobes accurately.

\begin{figure}
\centering
\includegraphics[angle=90,width=0.9\linewidth]{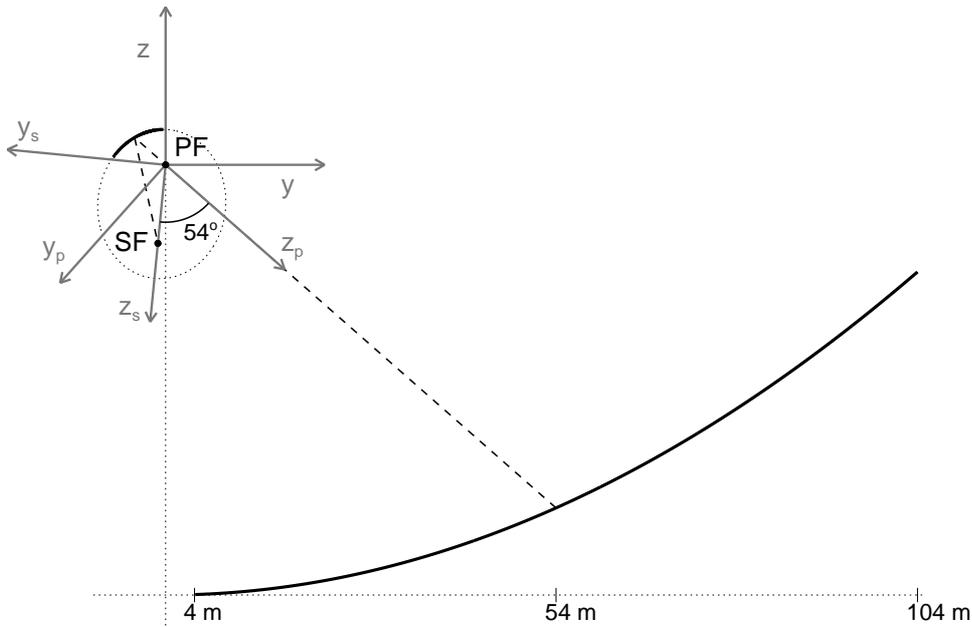}
\caption{Geometry of the GBT used in Antenna Workbench calculations.
SF and PF are the two foci of the ellipsoid of which the subreflector is
a segment.  PF is also the primary focus of the main parabolic
reflector.
The ``$s$'' coordinate system used for the near-field subreflector
patterns is aligned with the $z_s$ axis running along the ellipsoid axis
from PF toward SF.
The illumination of the primary by these patterns is given in the
``$p$'' coordinate system, rotated with respect to ``$s$'', so that the
$z_p$ axis runs from PF toward the projected center of the primary.
The all-sky far-field response is given in the non-subscripted
coordinate system, where the $z$ axis runs from PF in the direction of
the main beam.
All have $y$ and $z$ in the symmetric plane of the antenna.  }
\label{GBT_schematic}
\end{figure}

Fig.~\ref{GBT_schematic} shows the geometry of the primary and
secondary reflectors of the GBT in the antenna symmetric plane that is
vertical given the Alt-Az mounting.
Three coordinate systems to be described further below are indicated.
The non-subscripted one is associated with the primary, for the
all-sky power pattern.  That associated with the subreflector is
subscripted ``$s$''.  A third coordinate system subscripted ``$p$'' is
used for subreflector illumination of the primary, including for
calculations of spillover past the primary.
These are right-handed $x-y-z$ systems, with the $x$ axis directed up
out of the page and $yz$ in the antenna symmetric plane.  In these
coordinate systems we define $\phi$ to be the azimuthal angle (range
$-90\degr$ to $90\degr$) around the $z$ axis, measured relative to the
$x$ axis ($\phi = 0\degr$) toward the $y$ axis ($\phi = 90\degr$).
The polar angle $\theta$ is measured in half-planes of constant $\phi$
with an edge along $z$, relative to the $z$ axis ($\theta = 0\degr$)
and ranging over $180\degr$ to the $-z$ axis.

The radiation pattern of the GBT was computed in two steps.  First,
the pattern of the subreflector was calculated, in the ``$s$''
coordinate system whose origin is at one of the foci of the ellipsoid
of which the subreflector is a segment.  This focus is also the
primary focus (PF) of the main reflector (Fig.~\ref{GBT_schematic})
and is the phase reference point for the subreflector patterns.  The
L-band feed horn is at the other focus of the ellipsoid, which is the
GBT Gregorian (secondary) focus (SF).  The $z_s$ axis runs from PF
toward SF.  These calculations used the measured far-field patterns of
the L-band feed horn \citep{Srikanth1993} to illuminate the
subreflector.  The grids on the subreflector were set at $5 \times
5$~cm.  Since the main reflector is in the near field of the
subreflector, the near-field option in the code was used.

\begin{figure}
\centering
\includegraphics[angle=90,width=1.\linewidth]{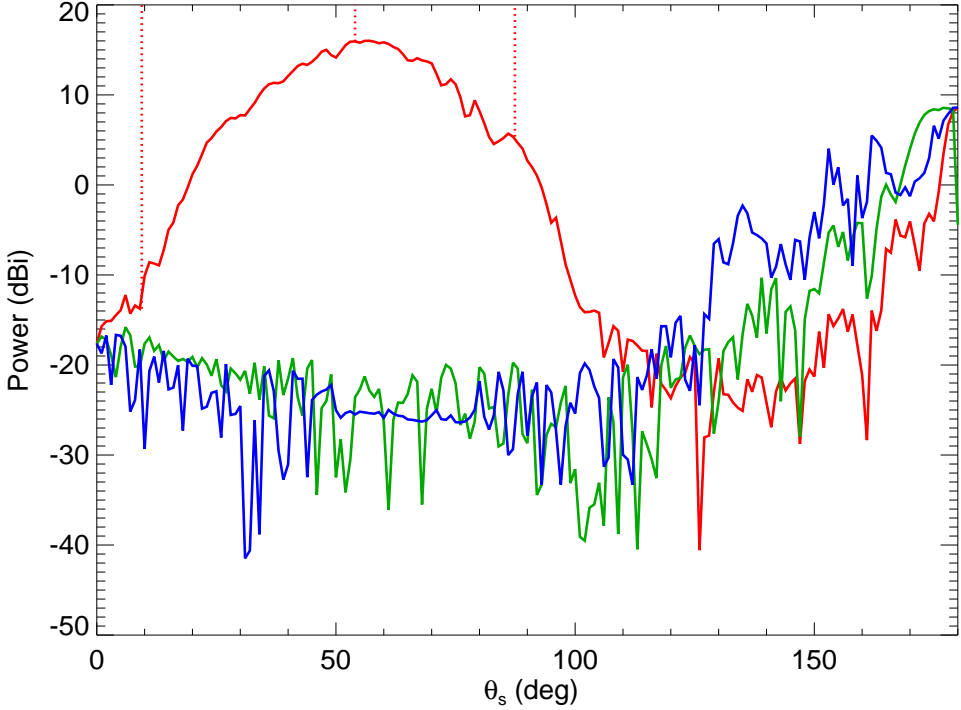}
\caption{Subreflector near-field patterns in the ``$s$'' coordinate
  system as a function of angle $\theta_s$ from the ellipsoid axis
  $z_s$, in half-planes with different values of $\phi_s$: 
$\phi_s = -90\degr$, red; %dark dot-dashed; 
$0\degr$, green; %light solid; 
$90\degr$, blue. %dark dashed.
Vertical lines indicate the near edge ($9\fdg4$), center ($54\fdg0$),
and far edge ($87\fdg4$) of the main reflector in the $\phi = -90\degr$
half-plane.  }
\label{subreflector_patterns}
\end{figure}  

Figure~\ref{subreflector_patterns} shows the power patterns calculated
for the subreflector in five $\phi_s$ half-planes.
Given the orientation of the ``$s$'' coordinate system, it is the
patterns in the half-planes approaching $\phi_s = -90\degr$ that
illuminate the main reflector.  In this coordinate system, in the
antenna symmetry plane the near edge of the main reflector is at
$\theta_s = 9\fdg4$ and the far edge is at $87\fdg4$.
At $54\degr$ is the center of the main reflector.\footnote{This is the
  projected center as seen along the direction of the main beam, $z$;
  the projected aperture diameter is 100~m.}
The power pattern shown for $\phi_s = -90\degr$ is appropriately
peaked to provide good illumination.

\begin{figure}
\centering
\includegraphics[angle=90,width=1.\linewidth]{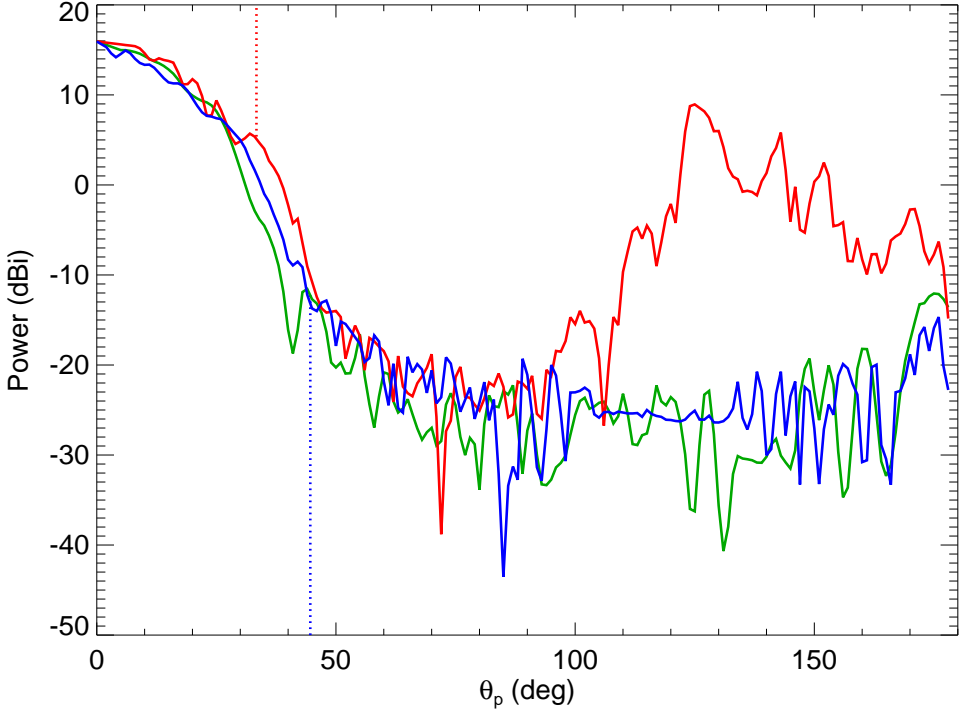}
\caption{Subreflector illumination patterns on the main reflector as a
  function of angle $\theta_p$ from $z_p$, the direction from PF to the
  center of the primary, for three values of $\phi_p$ in the main
  reflector (``p'') coordinate system: 
$\phi_p = -90\degr$, red; %dark dot-dashed;
$0\degr$, green; %light solid;
$90\degr$, blue. %dark dashed.  
Vertical lines indicate the near ($44\fdg6$) and far edges ($33\fdg4$)
of the main reflector in the $\phi = \pm 90\degr$ half-planes,
respectively.  }
\label{Sri_appendix_fig3}
\end{figure}

To calculate the illumination of the main reflector and the spillover
past its edge, these patterns were transformed from the ``$s$'' to the
``$p$'' coordinate system with origin at PF and axis $z_p$ oriented
from PF to the projected center of the 100-m main reflector
(Fig.~\ref{GBT_schematic}).
Fig.~\ref{Sri_appendix_fig3} shows the illumination of the main
reflector from the subreflector transformed to this coordinate system,
for three different $\phi_p$ half-planes.  The edge of the main
reflector subtends an angle that varies between $\theta_p = 33\fdg4$
and $44\fdg6$ as $\phi_p$ changes from $-90\degr$ (far edge) to
$90\degr$ (near edge).
As can be seen from both Figs.~\ref{subreflector_patterns} and
\ref{Sri_appendix_fig3}, the illumination taper in the antenna
symmetry plane at the far edge of the main reflector is about 5~dBi
and somewhat lower, $-13$~dBi, at the near edge, so that the spillover
lobe from the main reflector is not symmetrical around $z_p$ (see
below).

The second step was to determine the all-sky response of the GBT,
expressed in the non-subscripted coordinate system in
Fig.~\ref{GBT_schematic}.  Subreflector patterns as calculated by
Antenna Workbench for 72 half-planes in the range of $-90\degr \leq
\phi_s \leq 90\degr$ were used.  The main reflector was gridded into
regions $30 \times 30$ cm, which is about $1.4\lambda$, adequate for
the present work using AI. Far-field patterns were calculated at
$0\fdg05$ intervals in $\phi$ and $0\fdg02$ intervals in $\theta$.
The computation of this pattern switches from AI to GTD at $\theta =
2\fdg65 = \sin^{-1}(1/\sqrt{A_w})$ where $A_w$ is the aperture
diameter in wavelengths in any given $\phi$ plane.

\begin{figure}
\centering
\includegraphics[angle=90,width=1.\linewidth]{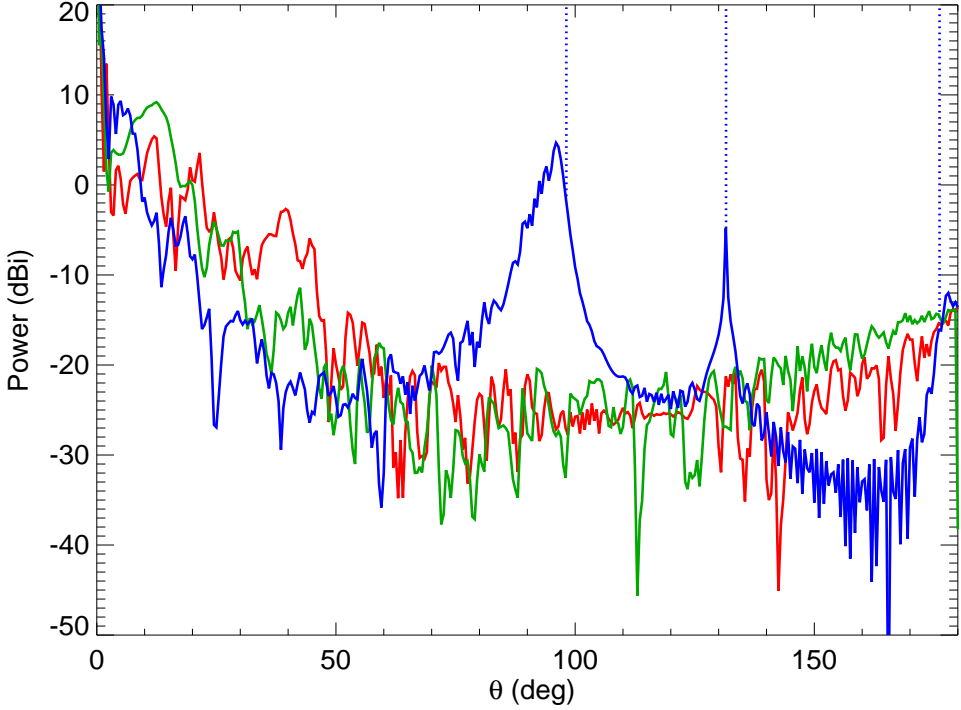}
\caption{Calculated all-sky response of the GBT at 1.4 GHz, the main
  beam and far sidelobes, as a function of angle from the main beam,
  $\theta$, for three values of $\phi$ in this coordinate system:
$\phi = -90\degr$ (towards the feed arm in the antenna symmetry plane),
  red; %dot-dashed; 
$0\degr$ (perpendicular to the antenna symmetry plane), green; %solid; 
$90\degr$ (away from the feed arm), blue. %dark dashed.
To show details of the sidelobes the scale has been magnified so that
the forward gain of the main beam, 61.6~dBi at $\theta = 0\degr$
(Fig.~\ref{mainbeam}, Table~\ref{eta_a_table}), is beyond the range of
this figure.
Vertical lines indicate the far edge ($98\fdg2$), center ($131\fdg5$),
and near edge ($176\fdg2$) of the main reflector in the $\phi = 90\degr$
half-plane.}
\label{Sri_appendix_fig4}
\end{figure}

Fig.~\ref{Sri_appendix_fig4} shows the main beam and the far sidelobes
of the GBT calculated for three different $\phi$ half-planes.  The
on-axis (forward) gain is 61.61~dBi.  An expanded view of the response
near the main beam is given in Fig.~\ref{mainbeam}.  The calculated
half-power beam width at 1.40~GHz is $9\farcm17 \times 9\farcm14$,
which differs from the measured values by $<1\%$ and $<2\%$,
respectively.

The sidelobes stay mostly below -20~dBi.  The peaks at about 10~dBi
seen for $\phi = 90\degr$ at 5\degree\ and for $\phi = 0\degr$ at
13\degree\ are the spillover sidelobe from the subreflector (Appendix
\ref{main}).  This can be appreciated in
Figs.~\ref{mainbeam_and_sidelobe_contour} and \ref{sunscan-sidelobes},
noting that the $-V$ axis there corresponds to $\phi = 90\degr$ and
the $H$ axis to $\phi = 0\degr$.

The $\phi = 90\degr$ plane is in the symmetric plane of the GBT
structure in the direction away from the feed arm and so as $\theta$
increases the main reflector is crossed.
The two sidelobe peaks near $96\degr$ and $178\degr$ are caused by the
spillover of the subreflector illumination
(Fig.~\ref{Sri_appendix_fig3}) past the far edge ($98\fdg2$) and near
edge ($176\fdg2$) of the main reflector, respectively.  Note that the
former is stronger because as shown above the illumination from the
subreflector is larger at the far edge.  This spillover lobe is
manifested as an asymmetrical ring around $z_p$ as $\phi_p$ changes.
The peak near $\theta = 131\degr$, which lies close to $z_p$
($131\fdg5$), is caused by the diffracted rays off the main reflector
edge, similar to the backlobe that is observed on an on-axis antenna.
This is like the Arago spot from the subreflector discussed in
Appendix~\ref{aragospot}.  Note that it is offset from the cone axis
defining the main reflector ($137\fdg2$) to lower $\theta$, toward the
far edge, because of the tilt of the reflector as seen from PF.

\section{Calculation of diffraction from the subreflector using
  Fresnel diffraction theory}
\label{secondary}

The primary reflector of the GBT is an offset segment of a symmetric
parabola designed to eliminate aperture blockage and attendant
sidelobes \citep{Prestage09}.  The Gregorian secondary subreflector is
an offset section of an ellipsoid with the prime and secondary foci of
the GBT at its two foci (Fig.~\ref{GBT_schematic}).  The GBT L-band
receiver feed horn is at the secondary focus.  The geometric path from
feed to subreflector to main reflector is clear of any obstruction.
However, to a receiver feed that does not have a perfectly sharp
cutoff of its field pattern at the subreflector edge, the subreflector
appears as a blockage of the sky and so its diffractive effects must
be considered.  The GBT L-band feed does indeed somewhat
over-illuminate (``spills over the edge of'') the subreflector,
because a feed having a more sharply tapered beam pattern would be too
large and heavy for the feed turret \citep{Norrod1996}.  Therefore the
field pattern of the feed couples to the diffracted radiation,
producing a modified Fresnel diffraction pattern, whose main ``edge''
feature is a set of annular peaks (rings) centered on the subreflector
and beyond its projected edge.  The dominant first ring is often
called the ``spillover sidelobe.''  Here we present a calculation of
the full power pattern.

\subsection{Simple diffraction pattern of the subreflector} \label{classical}

The shape of the subreflector is defined by the intersection of a
cone, with apex at the secondary focus and an opening half-angle
$\theta_H = 14.99 $\degree, with an offset ellipsoid of revolution (a
prolate spheroid).
As seen from the feed, the subreflector is to a first approximation a
circular blockage defined by radius $a = 3.775$~m at distance $d =
14.1$~m from the secondary focus.  At the operating wavelength of
21~cm, the secondary focus is in the near-field ($d < a^2/4 \lambda$)
with respect to the subreflector and therefore satisfies the Fresnel
condition.  Furthermore, $a^4/d^38\lambda < 1$ so that the usual
quadratic expansion of the phase used in the ``Fresnel approximation''
is valid (e.g., \citealp{shep92,daug96}).  Thus the characteristics of
the sidelobe pattern are closely related to the familiar near-field
Fresnel diffraction by a circular blockage.  This can be calculated
for a receiving system with a plane wave arriving along the cone
(negative $z_c$) axis or, by reciprocity, for a transmitting system
with an isotropic source at the feed.
We have evaluated this from the analytic solution for a receiving
system by \cite{hove89} and obtained the same results numerically
using the phasor approach described by \cite{daug96}, in which
reciprocity is implicit in the calculation.  Dauger's Fresnel
Diffraction Explorer (FDE)\footnote{http://daugerresearch.com/fresnel}
can also be used to explore this interactively.

\begin{figure} %[t]
\centering
\includegraphics[width=0.7\linewidth, angle=90]{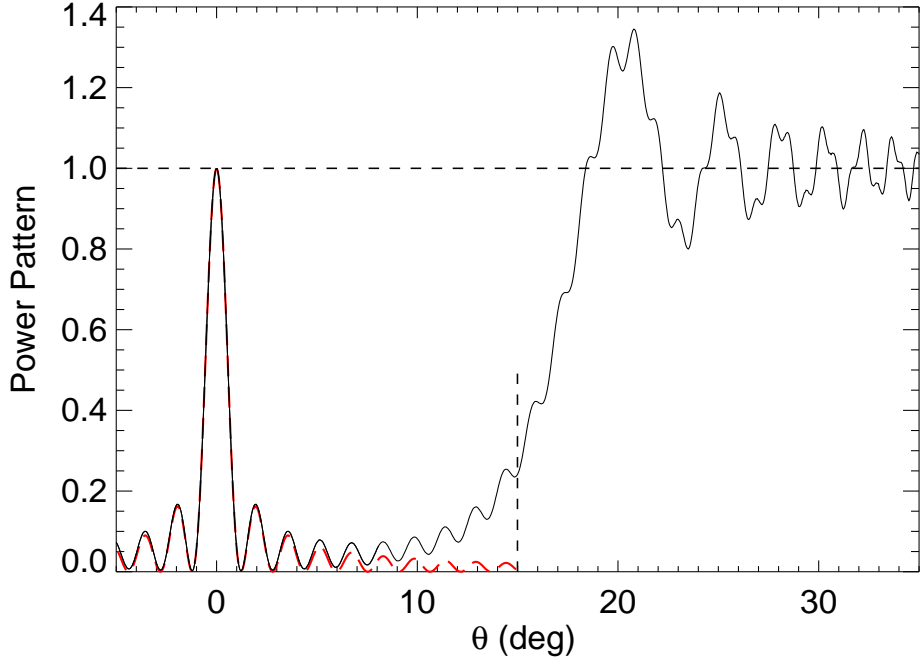}
\caption{Radial profile of the familiar diffraction power pattern of a
  circular blockage, $\theta$ being with respect to the axis of the cone
  defining the subreflector.  The vertical dashed line marks the edge of
  the subreflector as seen from the secondary focus.  The dashed curve
  approximating the pattern near the central Arago spot is the
  lowest-order term in the analytic solution by \cite{hove89}. }
\label{fresnel}
\end{figure}

A radial profile of the axially-symmetric power pattern is shown in
Fig.~\ref{fresnel}.  In this pattern, and in the GBT sidelobe pattern,
there are two main phenomena of interest: (i) the finely-spaced ring
structure near the cone axis with related modulations continuing at
angles beyond the edge of the subreflector, and (ii) the coarser ring
structure existing only beyond the edge. The existence of the
diffraction spot on axis depends on interference of waves arising from
all around the circumference of the subreflector in this
circularly-symmetric geometry.  On the other hand, the outer coarse
ring structure is an ``edge effect,'' depending only locally on the
sharp boundary, and so is qualitatively similar for a circular,
square, or knife-edge configuration.
As discussed below, the actual GBT sidelobe pattern shows the effects
of three other factors:
(i) the tapered beam of the L-band feed at the secondary focus,
(ii) a screen (Fig.~\ref{GBT_sub_screen}) adjacent to the subreflector
edge in front of the supporting feed arm to redirect feed spillover,
and 
(iii) the tilt of the subreflector.

\subsection{The Arago spot} \label{aragospot}

In the center of the diffraction pattern, at $\theta=0$, is the Arago
(or Poisson) spot, surrounded by a series of fine rings of comparable
width.  This pattern is described accurately by the lowest-order term
in the solution by \cite{hove89}, i.e., the square of the zero-order
Bessel function of the first kind, $J^2_0(v)$, where $v = k a r / d$,
$k = 2\pi/\lambda$, $r$ is the radial coordinate in the image plane
and $\theta = \arctan(r/d)$.  For the inner rings, the positions of
the minima and maxima are near the zeros and peaks of the function $1
+ \sin(2v)$, and thus at $(4n + zp)(\pi/4)/ka$, $n = 1, 2, ...$ and
$zp = -1$ or 1, respectively.  The first maximum is at $1\fdg99$.  In
these units, the half-power point of the main peak, at $v = 1.1263$,
is at angle $1.434(\pi/4)/ka$, and so the FWHM of the Arago spot is
$1\fdg14$; this is inversely proportional to $a$ and, for fixed cone
angle, also $d$.  The predicted FWHM and ring positions are in close
agreement with the properties of the actual scans of the GBT sidelobe
using the Sun.

\subsection{The main spillover sidelobe} \label{main}

\begin{figure} %[t]
\centering
\includegraphics[width=0.545\linewidth, angle=90]{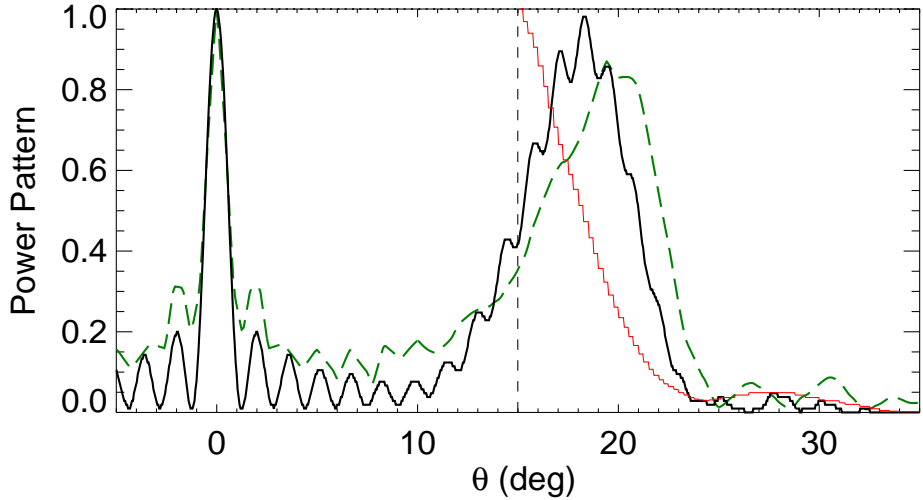}
\caption{
Solid (black): simulation of the effect of the tapered beam on the
diffraction pattern, normalized to unity at the Arago spot.
Step function (red): tapered field pattern of the L-band feed, relative
to unity at the edge of the subreflector.
Dashed (green): radial profile in the H direction through the observed
sidelobe pattern.
}
\label{taper}
\end{figure}

The dashed curve in Fig.~\ref{taper} shows the actual radial profile
of the observed sidelobe through the Arago spot along the H direction.
The oscillating coarse ring structure beyond the subreflector is an
edge effect.  However, the angular extent and amplitude are not the
same as the pattern shown in Fig.~\ref{fresnel}, because the strongly
tapered feed pattern, already down 14.7~dB at the edge of the
subreflector, does not illuminate the region beyond the edge of the
subreflector uniformly.  Thus the contribution to the diffraction
pattern from radiation at large angles is strongly suppressed.  We
have simulated this as a steadily increasing ``blockage'' beyond the
edge of the subreflector by implementing a method for treating
variable transmission suggested by \citet{daug96}.  The amplitude and
phase of the far-field pattern have been measured for the prototype
L-band feed \citep{Srikanth1993} and as assumed in
Appendix~\ref{calcSri} should be approximately valid for the distance
of the subreflector. The phase is roughly constant and we have taken
the transmission (coupling) to decrease following the field pattern.
The resulting diffraction pattern is shown in Fig.~\ref{taper},
normalized to the amplitude of the Arago spot.  The outer sidelobe
pattern is systematically and differentially diminished: the peak
intensity of the major spillover lobe compared to the Arago spot is
lower and the outer rings are suppressed even more.  As expected, the
position of the peak of the major ring moves closer to the edge of the
subreflector.  These effects are evident in the observed sidelobe
pattern, although the details are somewhat different; this might arise
in part because of the large angles encountered in the actual GBT
configuration.

\subsection{Effects of the screen} \label{screen}

The diffraction pattern is altered by the screen
(Fig.~\ref{GBT_sub_screen}) because it eliminates the contributions of
interfering waves from a wedge with angular range $\phi_s \approx
40$\degree\ along the circumference of the subreflector centered on
the positive V axis (away from the main beam).  The amplitude of the
Arago spot is reduced by a factor $(1. - \phi_s/360)^2 \approx 0.8$.
In the simulation for Fig.~\ref{taper} we have not included the
screen; this would lower the Arago spot while leaving the major
spillover lobe unaffected along the H profile thus bringing the
relative deflections even closer to that observed.

Our calculations incorporating the screen indicate that in the profile
along H the first fine ring is enhanced but in the orthogonal
direction, along V, the ring amplitudes are reduced while the radial
offset of the rings increases. These more subtle effects are
discernable in the observations.

Because the spillover sidelobe is an edge effect, the disruption of
the edge of the subreflector by the screen has a profound effect
locally.  This is clear in the observed pattern (see
Figs.~\ref{mainbeam_and_sidelobe_contour} and \ref{sunscan-sidelobes})
where there is a bite out of the spillover ring pattern, about a
40\degree\ wedge out from the Arago spot at positive $V$ centered on
the feed leg.  The screen apparently acts as an edge too, brightening
the adjacent portion of the spillover sidelobe.

Superimposed on the inner edge of the observed main spillover lobe at
negative $V$ near $H = 0$, on the side of the main beam opposite to
the direction of the screen, there is a pair of features in excess of
the simple diffraction pattern (see
Figs.~\ref{mainbeam_and_sidelobe_contour} and
\ref{sunscan-Arago-sidelobes}).  These have about the same beam
integral as the portion of the major spillover lobe removed in the
wedge, qualitatively consistent with conservation of energy.  The
screen has been well designed so that these features too lie well away
from the main beam.

\subsection{Effect of the tilt and motion of the subreflector} \label{tilt}

The required GBT geometry \citep{Norrod1996} is such that the
subreflector is not quite perpendicular to the axis of the cone, but
rather tilted by $t = $\tilt; relative to the defining cone axis, the
most distant edge is toward the main reflector and the direction of
the main beam (Fig.~\ref{GBT_schematic}).  The semi-major axis, in the
antenna symmetric plane and joining the edges of the subreflector
along the direction of tilt, is of dimension 3.975~m, and the
semi-minor axis is $a = 3.775$~m.  The distance along the cone axis
from the secondary focus to the center of the minor axis is then $d =
14.1$~m.  These were the values of $a$ and $d$ used above.
The tilt shifts the position of the Arago peak relative to the
direction to the subreflector center, such that there is still
constructive interference between waves from the distant and near
edges. Solving Eq.~A7 of \citet{daug96}, phases for the ends of the
major axis are equalized for an offset $2\fdg05 \tan t$ in the
direction toward the main beam.  For the specified tilt, our estimate
of the offset is $0\fdg63$.  The angle between the nominal cone axis
defining the subreflector center and the main beam is $12\fdg329$
\citep{Norrod1996} at the rigging angle, elevation \rigging\
\citep{Nikolic2007}, the geometry used in the above calculations.
Thus the expected offset of the Arago spot at the rigging angle would
be $11\fdg7$ along the V axis above the main beam.

At the rigging angle the main reflector is a perfect paraboloid to
21-cm radiation and the feed and subreflector are located optimally.
At any other elevation angle, the main reflector deforms, displaces,
and rotates while the feed and subreflector displace laterally in the
symmetry plane, resulting in a slight misalignment of the optics
\citep{sri1994}.
To track the primary focus position and maintain optimal efficiency
together with precision pointing, the subreflector is moved to
compensate (see, e.g., GBT Commissioning Memos 7 and
11\footnote{http://www.gb.nrao.edu/~rmaddale/GBT/Commissioning/memolist.html}).
Thus the angle between the cone axis of the subreflector and the main
beam changes slightly while tracking a source during an observation.

The elevation of the telescope when the Arago spot crossed the Sun
during two different nearly vertical scans was $31\fdg5$ and
$35\fdg4$. Allowing for the change with elevation we estimated that
the position of the Arago spot would be at $11\fdg8$.  The two scans
gave identical observed offsets, $11\fdg9$.  The good agreement is
perhaps fortuitous, given the complexity of the calculation.

\subsection{Comments on the derived sidelobe model} \label{diffraction}

These simple simulations reveal potential subtle issues regarding our
method of deriving the sidelobe pattern by scanning it with the Sun.
One involves the data reduction, where of necessity the scans have had
a baseline removed.  This introduces some uncertainty in the derived
pattern.
For example, in the area of the Arago spot there is a plateau not
explained by the above simple diffraction theory (Fig.~\ref{taper}).
Even though this plateau is relatively weak, its subtended area could
potentially lead to a significant contribution to the stray radiation.
Another possibility, with an effect of the opposite sign, is that the
derived pattern does not sink quite to zero in between the main
spillover lobe and next minor rings. Again, because of the large area,
even weak perturbations like this could have considerable power.  To
compensate for inadequate baseline removal, an approach could be to
introduce a smooth parameterized sidelobe contribution centered on the
subreflector and a function only of $\theta$, e.g., a low order
polynomial plus exponential.  The parameters could be optimized during
the procedure described in \S~\ref{getfsun} to determine the optimal
scaling of the sidelobe pattern.  The effective coverage of different
parts of the sky by the sidelobe pattern would therefore be
reweighted.

A second issue concerns how the actual diffraction pattern might
change in position with elevation of the telescope away from the
rigging angle.  As discussed in connection with the position of the
Arago spot, displacement of the cone axis of the subreflector with
respect to the main beam is difficult to calculate precisely but it
might be several tenths of a degree over the range of elevations used
in our survey, from $15\degr$ almost to the zenith.  A smaller range
was used in our mapping of the sidelobe pattern with the Sun.  Because
major features of the sidelobe pattern are related to subreflector
spillover, and the latter is geometrically tied to the cone axis, the
whole pattern can shift slightly with elevation.  As a further
complication, the screen is fixed to the feed arm, not moving with the
subreflector during focus tracking.  Thus the sidelobe pattern as
measured and implemented will not be perfect; this might account (at 
least in part) for the slight differences among the optimal values 
of $\eta_{\rm sl}$ derived independently in different survey regions
(Sect.~\ref{sunscanscale}) and for the stray radiation correction errors
--- which are nevertheless small (Sects.~\ref{neptest} and
\ref{S6S8calib}).

\section{Stray radiation correction algorithm} \label{program}

The stray radiation correction is computed by evaluating the integral
in Eq.~\ref{removestray}.  This is based on a program written in 1997
by Edward M. Murphy to correct stray radiation for \HI\ spectra from
the 43~m (140~ft) telescope at the NRAO incorporating a
multi-component semi-analytic model developed from careful
measurements of the 43~m telescope and its sidelobes (E.M. Murphy,
1993, unpublished M.A. thesis, University of Virginia).  For each
input \HI\ spectrum, an output spectrum of the estimated stray
radiation was produced, with the input file serving as a template for
the output file.  The algorithm to estimate the sidelobe was
straightforward, using an empirical model of the \HI\ sky augmenting
an early version of the Leiden-Dwingeloo survey with data from the
Bell Labs survey and the Parkes 60-foot survey
\citep{1986KerrA,1992Stark,1997Hartmann}.  This model sky had \HI\
spectra at half-degree intervals in Galactic latitude~$l$ and
longitude~$b$.  For each $(l,b)$ point in the model, the date and time
of the input spectrum were used to determine the corresponding
position on the sky as viewed from the telescope.  If this point lay
above the horizon (it is assumed that there is no \HI\ emission
reflected from the ground; see Appendix~\ref{stray_algorithm_errors})
and was more than one degree from the main beam, then the
corresponding survey spectrum was multiplied by the relevant solid
angle, by a factor to account for atmospheric absorption at the
corresponding zenith angle, and by the sidelobe level at that position
relative to the main beam.  The appropriate velocity shift relative to
the LSR was applied and the spectrum was spline-interpolated onto the
input-spectrum velocity grid.  The sum of all such weighted
velocity-shifted spectra, looping over the grid of the model \HI\ sky,
was output as the stray radiation estimate.  This computation of one
stray-radiation spectrum required about one minute of CPU-time on a
typical workstation.

\subsection{Modified algorithm for the GBT} \label{updated-program}

Major modifications to support calculation of stray radiation
corrections specifically for the GBT involved four main areas: adding
support for the SDFITS file format used for GBT observations,
modifying the sky position algorithms to account for the fact that the
GBT has an altitude-azimuth mount instead of the equatorial mount of
the 43~m telescope (as well as including the horizon as seen from the
GBT), modifications to speed up the calculation by a couple of orders
of magnitude, and finally obtaining a good map of the GBT sidelobes
(Sect.~\ref{sidelobes}) which is stored in and interpolated from a 2-D
array.

For the empirical model \HI\ sky we adopted the LAB all-sky survey
\citep{Kalberla2005} after first subtracting a Gaussian with peak
amplitude $0.048$~K, center at $-22$~\kms, and FWHM${} = 167$~\kms\
from each LAB spectrum (the so-called ``Wakker correction'',
\citealp{2011Wakker}) and then dividing by a scale factor \labvsgbtratio, as
discussed in Sect.~\ref{LABscale} (see Eq.~\ref{lab_match_zero_a}).
The ``Wakker correction'' to the LAB survey has a very small effect on
the calibrated GBT spectra as it enters scaled by $\eta_{\rm sl}$
times the fraction of $\eta_{\rm sl}$ that is above the horizon.  As
the product of these quantities is always $<0.1$, its effect on
calibrated GBT spectra is always $<0.005$~K.

Air mass measurements were available for NRAO at 5\degr\ intervals in
elevation angle~$el$ for $0\degr \le el \le 90\degr$ and at 1{\fdg}5
intervals for $0\degr \le el \le 15\degr$ (R.D. Maddalena 2005,
private communication: the air mass measurements were actually
performed nearby at Hot Springs, VA, on 15 June 2005).  These
measurements were used to obtain the atmospheric optical
depth~$\tau_{\rm atm}$ as a function of~$el$: a zenith optical depth
at 1420~MHz of $\tau_{\rm zenith} \approx 0.01$, and an air mass of
roughly~31 at the horizon, implies that $\tau_{\rm atm} \approx 0.32$
at $el = 0\degr$ (rather than becoming infinite there).  However, as
the sky at $el < 1\degr$ is obscured by hills around most of the
horizon, the improved atmospheric absorption values had little effect
on the calculated stray radiation.  Information on the exact weather
conditions is not used at present when calculating the opacity, though
there are indications in our data that rain can add an extra opacity
that might be accounted for in the future (Sect.~\ref{S6S8calib};
R.D. Maddalena, private communication).  Snow might also have a
significant effect both with its precipitation and with its
accumulation in the GBT dish --- any effects from the latter not being
sufficiently predictable to be corrected for.  However, variations in
atmospheric pressure, temperature, and humidity are expected to have
very little effect on the atmospheric absorption; this is confirmed by
the fact that the repeated observations of S6 and S8
(Sect.~\ref{S6S8calib}) show no evidence for a systematic
summer-vs.-winter effect.

Refraction results in more of the sky being visible to the telescope,
an extra strip approximately $0{\fdg}3$ wide around the horizon.
Effects of temperature, pressure, and humidity {\em are\/} included in
the refraction calculations.  Inclusion of refraction typically
changes the calculated stray radiation by less than 1\%, but up to
10\% in extreme cases.

For integration times of a few minutes or more, there are significant
changes in the position of the sky relative to the horizon and, for
the GBT, in the rotation of the sidelobe beam relative to the sky with
accompanying changes in the Doppler shift of the stray radiation being
accumulated.  For an integration time of 120~sec, such as we used for
``staring observations'' as opposed to on-the-fly mapping, calculating
the sky, horizon, and beam positions at the middle of the time
interval (rather than the beginning) yields typical improvements of a
fraction of a percent though in extreme cases it can be 10\%.  To
allow for even longer integration times, calculation of stray
radiation contributions at multiple times within a single observation
is supported.

The calculation was sped up mainly by two modifications.  First,
speed-up by nearly an order of magnitude was obtained simply by using
linear rather than spline interpolation in velocity for the spectrum
being accumulated.  Second, another order of magnitude was obtained by
introducing a four-level tiling tree for the model sky spectra: a
four-level tesselation of the sky.
The deepest level corresponds to independent spectra in roughly
$0{\fdg}5 \times 0{\fdg}5$ areas on the sky, appropriate to the LAB
survey sampling.  Near the Galactic equator, these tiles are at
0{\fdg}5 intervals in~$b$ and $l$, i.e., the LAB grid. Near the
Galactic poles, LAB spectra at multiple $l$~values were averaged
together to create tiles with widths {\em on the sky\/} of at
least~0{\fdg}25 but no more than~0{\fdg}5.
The next level up, coarser tiles of roughly $1\degr \times 1\degr$,
comprised of averages (weighted by solid angle) of four (or, near the
poles, sometimes three) of the lowest-level spectra.
Similarly, two further higher tiling levels of roughly $2\degr \times
2\degr$ and $4\degr \times 4\degr$ were obtained.
The loop over the tiles of the model starts at the highest $4\degr
\times 4\degr$ level. Since much of the sidelobe area is quite smooth,
this tiling made it possible to reduce greatly the number of spectra
that had to be accumulated for each stray calculation.  Where the
sidelobe varied significantly on length scales smaller than the
relevant tile, or where part of the tile would be below the horizon or
adjacent to the main beam, the appropriate tiles on the levels below
were traversed instead.
The magnitude of the introduced errors was estimated by computing
stray radiation corrections for a random sample of positions, dates,
and LST values.  For these modifications that yield faster sidelobe
spectrum evaluation, the resulting errors are small (rms~0.2\%,
max~3\% of the stray correction).

\section{Uncertainties not related to the GBT response pattern}
 \label{ssec:other}

\subsection{Undersampling in the LAB survey}
\label{undersampling}

The correction for stray radiation relies on the LAB survey to provide
a 21 cm $T_{\rm b}$ over the entire sky visible from Green Bank. The
main contribution to the LAB survey for the part of the sky seen by
the GBT is the Leiden-Dwingeloo (LD) \HI\ survey that consists of
observations spaced $0\fdg5$ on the sky at an angular resolution of
$35\arcmin$ \citep{1997Hartmann}.  It is sometimes said that the
undersampling makes the LD survey equivalent to a survey with an
angular resolution of order 1\degr, but this is not correct.  Because
of the undersampling, the LD survey (and hence the LAB in the North)
cannot be used for accurate interpolation between the measured
positions -- information at spatial frequencies that are resolvable to
the Dwingeloo antenna are aliased into lower spatial frequencies
biasing our estimate of the \HI\ sky.  At present there is no remedy
for this and it contributes an unknown error to the stray radiation
correction.  A new \HI\ survey of the northern sky is underway that
should provide a much more accurate, complete, data set from which to
calculate the stray component in GBT spectra \citep{Kerp2011}.

\subsection{Errors in the LAB survey}
 \label{LAB_errors}

 Any errors in the LAB survey spectra will propagate through the stray
 radiation correction, reduced in their effect by $\eta_{\rm sl} =
 \slvalue$ and the fractional solid angle of the erroneous spectra,
 and by the fact that multiple errors in opposite directions might
 cancel out.  We consider here errors in the LAB survey discovered
 earlier \citep{Higgs2005} and during the course of this work.

As discussed in Sect.~\ref{LABscale}, a very few of the LAB survey
spectra have normalization errors of up to about 30\%, but only a
small fraction of the spectra appear to have normalization errors even
at the ten percent level.

As noted earlier by \citet{Higgs2005} when they compared their DRAO
26-m data to the LAB survey, a very few LAB spectra have large
spurious features.  These are detected by comparison of LAB with other
data; the fact that it is the LAB feature that is spurious is
indicated by the fact that (1)~the feature appears in the LAB spectrum
but not in the other data, (2)~nor does the feature appear in LAB
spectra at adjacent latitudes, and (3)~the feature often has a shape
that is not typical of any real feature.  (Note that LAB survey
spectra at adjacent {\em longitudes\/} are not independent at high
latitudes, where a spurious feature might thus propagate with varying
intensity across several longitude points.)  Only where alternative
data are available can the above errors be reliably detected and
corrected.  This alternative is available for only 13165 of the 257762
LAB spectra; in 125 of these 13165 spectra, a total of 141 spurious
features were found (including 33 in the regions surveyed by
\citealt{Higgs2005}).  Most were small, ``dips'' or ``peaks'' with a
width of about~2~\kms\ superimposed on real features an order of
magnitude larger in both width and height.  However, 6 of the LAB
spectra had very wide spurious ``peaks'' at negative velocities.  The
LAB spectrum at $(l=138\fdg5, b=37\degr)$ has a ``peak'' over the
range $-300\,\kms < v < -160\,\kms$, with a plateau of height 3.6~K in
the range $-240\,\kms < v < -190\,\kms$ (this region should be zero);
this feature also shows up at $l = 139\degr$ and $139\fdg5$, albeit
with lower amplitudes of 0.5 and 0.1~K, respectively.  The LAB
spectrum at $(l=141\degr, b=41\fdg5)$ has a peak of height 0.6~K in
$-190\,\kms < v < -70\,\kms$ (this region should be zero).  Finally,
the two LAB spectra at $(l=132\degr$ and $132\fdg5, b=47\degr)$ have a
peak of height 2~K in $-240\,\kms < v < -20\,\kms$ (the convolved GBT
spectra, and adjacent LAB spectra, have three narrower peaks in this
velocity range, with heights of about~0.2~K).  Replacing the spurious
features by a linear interpolation over the relevant velocity range
works very well, except in the last two cases (where convolved GBT
spectrum values can be used instead, in the relevant velocity region).

The existence of spurious LAB features and random LAB calibration
errors in our large GBT fields allows for some quantification of the
potential propagated error.  For pointings near the middle of these
fields, the major contribution to the stray radiation comes from the
LAB spectra that were corrected for the above two types of errors, and
comparing these cases to computations without these corrections
indicates that these spurious features and mis-calibrations introduce
errors of at most a few percent in the computed stray radiation
correction.  Thus any such spurious features that cannot be corrected
(i.e., in regions with no comparison data) are not expected to
contribute significant errors to the stray radiation computation.

\subsection{Errors from the stray-correction algorithm}
 \label{stray_algorithm_errors}

The program currently assumes that no 21-cm radiation is picked up
through reflection from the ground; the tree-covered mountains around
the GBT can be assumed to be good absorbers, but the fields in the
immediate vicinity of the telescope might have a non-trivial albedo
especially when wet.  This was found to be the case for the moors
around Dwingeloo \citep{Hartmann1996}.  However, as the dominant GBT
sidelobes are near and above the main beam
(Fig.~\ref{mainbeam_and_sidelobe_contour}), the fraction of the
response lying on the ground is at most a few percent and so the
effect of reflections should be quite small.

The ``sidelobes'' over which the stray radiation is accumulated are
considered to begin a degree from the center of the beam.  Shifting
this cutoff from a $1.0\degr$ radius to $0.9\degr$ has an effect of
rms~2\% (max~14\%) on the stray radiation, according to a set of
random test cases.  Such effects are expected to be most significant
when the GBT beam is pointed at a region with extended strong signals,
such as the Galactic plane.  For pointings in regions of low signal,
even much larger changes in the definition of where the sidelobes
begin have little effect.

As discussed above (Sect.~\ref{diffraction}), variations in the
geometry of the optics with telescope elevation angle can shift the
sidelobe pattern by several tenths of a degree (its shape and
amplitude pattern might also change slightly).  This is at least an
order of magnitude smaller than the angular scale on which the
sidelobes vary, but could nonetheless lead to significant stray
radiation errors for cases where the mean sky brightness is varying
significantly at the edge(s) of the spillover lobe.  Typically, such
errors would be expected to be no more than a few percent of the
calculated stray radiation; but in extreme cases, this effect could be
larger.  The present stray-correction algorithm ignores this telescope
elevation effect, but in principal measurements might allow at least a
partial correction for it.

For pointings in bright complex regions, spectra obtained at different
times might change by a few percent due to the rotation relative to
the sky of the slightly asymmetric main beam and inner sidelobes
within $1.0\degr$ radius.  Such ``near-beam'' effects cannot be
corrected by this program due to the coarse grid of the all-sky LAB
survey.  This could be approached as a deconvolution problem using the
accurately known near beam if a small fully sampled map were obtained
quickly with the GBT\hbox{}.  Given our interest in high latitude
emission, this has not been pursued here.

\end{appendix}

\end{document}